\documentclass{aa}  

\usepackage{graphicx}
\usepackage{txfonts}
\usepackage{rotating}
\usepackage{answers}
\usepackage{xcolor}
\usepackage{scalefnt}
\usepackage{xcolor}
\usepackage{ragged2e}
\usepackage{float}

\usepackage[colorlinks=true, allcolors=blue]{hyperref}
\newbox\grsign \setbox\grsign=\hbox{$>$} \newdimen\grdimen \grdimen=\ht\grsign
\newbox\simlessbox \newbox\simgreatbox
\setbox\simgreatbox=\hbox{\raise.5ex\hbox{$>$}\llap
     {\lower.5ex\hbox{$\sim$}}}\ht1=\grdimen\dp1=0pt
\setbox\simlessbox=\hbox{\raise.5ex\hbox{$<$}\llap
     {\lower.5ex\hbox{$\sim$}}}\ht2=\grdimen\dp2=0pt

\begin{document}



\title{Combined Gemini-South and HST photometric analysis of the globular cluster NGC~6558 \thanks{Observations collected at the Gemini South Telescope, under programme GS-2020A-Q-121 (PI: L. Kerber), and at the \emph{Hubble} Space Telescope under programs GO-15065 (PI: R. Cohen) and GO-9799 (PI: Rich)}} 
\subtitle{The age of the metal-poor population of the Galactic Bulge}
   
   \author{S. O. Souza\inst{1,2}
          \and
          M. Libralato\inst{3,4}
          \and
        D. Nardiello\inst{3,5,6}
        \and
        L. O. Kerber\inst{7}
        \and
        S. Ortolani\inst{3,5,6}
        \and 
        A. P\'erez-Villegas\inst{8}
        \and
        R. A. P. Oliveira\inst{9}
        \and
        B. Barbuy\inst{1}
        \and
        E. Bica\inst{10}
        \and
        M. Griggio\inst{3,11}
        \and
        B. Dias\inst{12}
                  }

\institute{Universidade de S\~ao Paulo, IAG, Rua o Mat\~ao 1226,
Cidade Universit\'aria, S\~ao Paulo 05508-900, Brazil\\
              \email{stefano.souza@usp.br}
\and
Max Planck Institute for Astronomy, K\"onigstuhl 17, D-69117 Heidelberg, Germany \\ \email{s-souza@mpia.de}
\and
INAF-Osservatorio di Padova, Vicolo dell'Osservatorio 5, I-35122 Padova, Italy
\and
AURA for the European Space Agency (ESA), Space Telescope Science Institute, 3700 San Martin Drive, Baltimore, MD 21218, USA
\and
Universit\`a di Padova, Dipartimento di Astronomia, Vicolo dell'Osservatorio 2, I-35122 Padova, Italy
\and
Centro di Ateneo di Studi e Attivit\`a Spaziali “Giuseppe Colombo” – CISAS, Via Venezia 15, 35131 Padova, Italy
\and
Universidade Estadual de Santa Cruz, Departamento de Ci\^encias Exatas, Rod. Jorge Amado km 16, Ilh\'eus 45662-900, Bahia, Brazil
\and
Instituto de Astronom\'ia, Universidad Nacional Aut\'onoma de M\'exico, A. P. 106, C.P. 22800, Ensenada, B. C., M\'exico
\and
Astronomical Observatory, University of Warsaw, Al. Ujazdowskie 4, 00-478 Warszawa, Poland
\and
Universidade Federal do Rio Grande do Sul, Departamento de Astronomia, CP 15051, Porto Alegre 91501-970, Brazil
\and
Dipartimento di Fisica, Universit\`a di Ferrara, Via Giuseppe Saragat 1, Ferrara I-44122, Italy
\and
Universidad Andres Bello, Facultad de Ciencias Exactas, Departamento de Ciencias Físicas, Instituto de Astrofísica, Av. Fernández Concha 700, Santiago, Chile
}

\date{Received 20/05/2024 ; accepted 19/07/2024}

 
  \abstract
   {NGC~6558 is a low-galactic latitude globular cluster projected in the direction of the Galactic bulge. Due to high reddening, this region presents challenges in deriving accurate parameters, which require meticulous photometric analysis. We present a combined analysis of near-infrared and optical photometry from multi-epoch high-resolution images collected with Gemini-South/GSAOI+GeMS (in the $J$ and $K_S$ filters) and HST/ACS (in the F606W and F814W filters).}
   {We aim to refine the fundamental parameters of NGC~6558, utilizing high-quality Gemini-South/GSAOI and HST/ACS photometries. Additionally, we intend to investigate its role in the formation of the Galactic bulge.}
   {We performed a meticulous differential reddening correction to investigate the effect of contamination from Galactic bulge field stars. To derive the fundamental parameters age, distance, reddening, and total-to-selective coefficient, we employ a Bayesian isochrone fitting. The results from high-resolution spectroscopy and RR Lyrae stars were implemented as priors. For the orbital parameters, we employ a barred Galactic mass model. Furthermore, we analyze the age-metallicity relation to contextualize NGC~6558 within the Galactic bulge history.}
   {We studied the impact of two differential reddening corrections on the age derivation. When removing as much as possible the Galactic bulge field star contamination, the isochrone fitting combined with synthetic colour-magnitude diagrams gives a distance of $8.41^{+0.11}_{-0.10}$ kpc, an age of $13.0\pm 0.9$ Gyr, and a reddening of E($B-V$)$\,\,=0.34\pm0.02$. We derived a total-to-selective coefficient R$_V = 3.2\pm0.2$ thanks to the simultaneous near-infrared$-$Optical synthetic colour-magnitude diagram fitting, which, aside from errors, agrees with the commonly used value. The orbital parameters showed that NGC~6558 is confined whitin the inner Galaxy and it is not compatible with a bar-shape orbit, indicating that it is a bulge member. Assembling the old and moderately metal-poor ([Fe/H]$\,\,\sim-1.1$) clusters in the Galactic bulge, we derived their age-metallicity relation with star formation stars at $13.6\pm0.2$ Gyr and effective yields of $\rho=0.007\pm0.009\,\, Z_\odot$.}
   {The derived old age for NGC~6558 is compatible with other clusters with similar metallicity and a blue horizontal branch in the Galactic bulge, which compose the moderately metal-poor globular clusters. The age-metallicity relation shows that the starting age of star formation is compatible with the age of NGC~6558, and the chemical enrichment is ten times faster than the ex-situ globular clusters branch.}

    
   \titlerunning{NGC~6558 and MMP bulge GCs AMR}
   \authorrunning{Souza, S. O et al.}
   \maketitle
%
    
\section{Introduction}

{ The in situ Galactic bulge globular clusters (GCs) are very likely the first objects formed in the early Galaxy. They should have formed in the early building blocks that merged into a proto-Milky Way (MW) within the $\Lambda$CDM scenario \citep{whiterees78,whitefrenk91}. The important discriminator of the phase, when they formed, is their very old age $> 10$ Gyr \citep{kruijssen19,kerber19,ortolani19,forbes20,souza21,souza23}, older than the thick disk \citep[$\sim 6-9$ Gyr;][]{pinna24}  and bar formation \citep{barbuy18a}. The Galactic bar is supposed to have been formed 8 Gyr ago \citep{bovy19,wylie22} with evidence of a recent burst of star formation 3 Gyr ago \citep{nepal23}. }

The present composition of the Galactic bulge is far more complex  than was thought previously, with a mix of stellar populations \citep[e.g.][]{barbuy18a,queiroz20,queiroz21}. The present Galactic bulge includes the presence of a bar and a modified bulge or pseudo-bulge, inner thin and thick disc \citep{nogueraslara_disc}, inner halo \citep{perez_innerhalo}, a nuclear star cluster \citep{nogueraslara_nsc}, as well as accreted structures such as Heracles \citep{horta21}, Kraken \citep{kruijssen20}, and a smaller contribution from \emph{Gaia}–Enceladus–Sausage \citep{belokurov18,helmi18}.

{The GCs located in the inner Galaxy, as listed by \cite[][their Table 3]{bica16,bica24} and classified as Galactic bulge clusters based on orbital analysis \citep{perez-villegas20} are also identified as formed in-situ by \cite{belokurov24}. In particular,  it is not possible to define the most recently detected clusters \citep{bica24,garro24} as in-situ or ex-situ because there is not enough information to classify them (e.g., chemical abundances, ages, orbital properties).} {Regarding their orbit shapes, they would be modified by the gravitational force induced by the bar formation, with some of them trapped in one resonance of the bar. Simulations of the bar effects on the kinematics and structure of the Galactic box/peanut bulge can be found in \cite{debattista17}, \cite{fragkoudi20}, \cite{moreno22}, among others.}

{In the present work, we carry out a Colour-Magnitude Diagram (CMD) study of the GC NGC~6558. The cluster is projected on the Galactic bulge, with equatorial coordinates (J2000) $\alpha = 18^{\rm h}10^{\rm m}18.4^{\rm s}$, $\delta = -31^{\rm o}45'49''$, and Galactic coordinates $l = 0.201^\circ$, $b = +6.025^\circ$ \citep[2010 edition]{harris96}. 
{In \citet{rich98}, NGC~6558 was analysed and suggested to belong to a new (at that moment) class clusters in the Galactic bulge characterized by having a clear blue extended horizontal branch and a poorly populated giant branch. }
The cluster metallicity of $-1.17\pm0.10$ was recently derived from high-resolution spectroscopy \citep{barbuy18a,gonzalez23}. 
Photometric and spectroscopic analyses have shown that the Galactic bulge GCs have a peak in metallicity around [Fe/H]$=-1.1$ \citep[][]{bica16,bica24}, completing the characterisation of this class of clusters. So far, a dozen clusters have been identified with a metallicity of [Fe/H]$\sim-1.1$ together with a very old age and located in the Galactic bulge \citep[e.g.][]{garro23,bica24}. }

 \citet{perez18} integrated the NGC~6558 orbit, based on the \citet{rossi15} proper motions, combined with the radial velocity of $-194.45$ km s$^{-1}$ from \citet{barbuy18b},indicating that it shows a prograde orbit. The orbit also seems to follow the bar shape in the x–y projection and with a boxy shape in the x–z projection (in a bar co-rotating frame), indicating that NGC~6558 is trapped by the Galactic bar. Afterwards, \citet[][hereafter PV20]{perez-villegas20}, using \emph{Gaia} DR2 proper motions \citep{gaia18a}, computed the orbits of 78 Milky Way GCs listed in \citet[][their tables 1 and 2]{bica16}, which includes NGC~6558. Using a heliocentric distance of $8.26\pm0.53$ kpc \citep{barbuy18b} for NGC~6558, the cluster was identified as a current bulge/bar member with a $\sim 99 \%$ probability. 

{\cite{massari19}, using the \cite{mcmillan17} Galactic model, associated NGC~6558 with the main-bulge progenitor (in-situ) by means of the age-metallicity relation (AMR) and the total orbital energy and angular momentum in the z-direction. \cite{callingham22} have reached the same result through a robust statistical analysis of the AMR. More recently, \citet{belokurov24} classified NGC~6558 as an in-situ cluster following a classification method using the AMR and also detailed chemistry and dynamical properties.  }

It is important to stress the crucial role of distance in the orbital integration and, therefore, in the classification of NGC~6558 as a Galactic bulge member and in-situ cluster. In the literature, the heliocentric distance of NGC~6558 ranges from $\sim 6.3$ kpc \citep[][]{rich98} to $\sim8.3$ kpc \citep{barbuy18b}. Therefore, a comprehensive study of NGC~6558 supported by high-quality data is needed to consolidate the recent findings concerning its nature.

In the present work, we analyse NGC~6558 with deep images in the NIR obtained with Gemini-South/GSAOI and in the optical with ACS/HST. The simultaneous isochrone fitting in the two wavelength ranges allows better fixing the total-to-selective extinction ratio R$_V$. In Section 2 the observations and data reduction are described. In Section 3 the RR Lyrae are identified and used to derive the magnitude level to be used as a prior to the distance. In Section 4 the isochrone fitting is applied and described. In Section 5 the orbital analysis is presented. In Section 6 the results are discussed, and in Section 7 conclusions are drawn.



\section{Observations}

\begin{figure*}[hbt]
    \centering
    \includegraphics[width=0.8\textwidth]{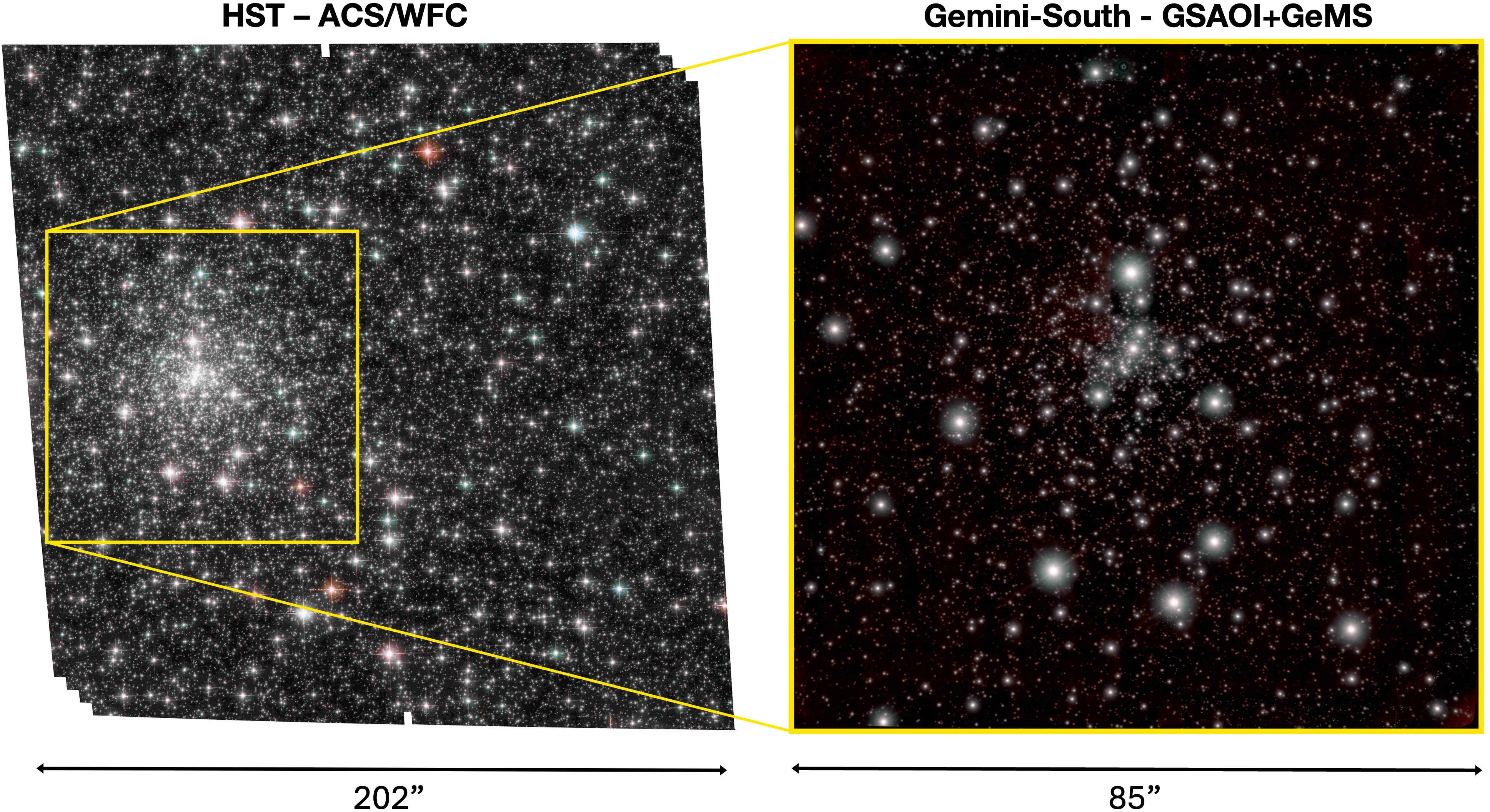}
    \caption{Images of NGC~6558. Left: $F606W/F814W$ combined colour image from the HST ACS/WFC camera for NGC~6558 ($202\times202$ arcsec$^2$). The green square represents the FoV of GSAOI ($85\times85$ arcsec$^2$). Right: $J/K_S$ combined colour image from the Gemini GSAOI camera. North is at the top, and east is to the left for both images.}
    \label{fig:hst_image}
\end{figure*}

The adopted HST data consists of two epochs of observations collected with the ACS/WFC imager: (i) GO-9799 data-set (PI: Rich) contains  1$\times$ 340~s F606W, 2$\times$10~s $+$ 1$\times$340~s F814W  images; (ii) GO-15065 data-set (PI: Cohen) contains 1$\times$10~s $+$ 4$\times$498~s F606W, 1$\times$10~s $+$ 4$\times$498~s F814W images. The difference between the first and the second epoch is $\Delta t \simeq 15.9 $~yrs. The HST image of the cluster combining the ACS/WFC filters used in this work, is shown in the left panel of Figure \ref{fig:hst_image}.

The GSAOI images of NGC~6558 were collected on 2020 March 3 as part of the program GS-2020A-Q-121 (PI: L. Kerber). They covered a field of view (FoV) of about 85 $\times$ 85 arcsec$^2$, with a pixel scale of 20 mas pixel$^{-1}$. The field centered on NGC~6558 was observed in $J$ and $K_{\rm{s}}$ filters with a dithering pattern of 5 images with 4 coadds of 50~s, totalizing 1000~s of exposure time in each filter. Additional $J$ and $K_{\rm{s}}$ images with an exposure time of 10~s were obtained to extend the analysis to bright stars otherwise saturated in longer exposures. The median FWHM value for the $J$ and $K_{\rm{s}}$ images is about 0.10 arcsec. The Gemini image of the cluster combining the GSAOI $J$ and $K_S$ filters used in this work, is shown in the right panel of Figure \ref{fig:hst_image}.

\subsection{Data reduction and proper-motion computation}

%
%
The GSAOI data reduction was performed following the procedure described in \citet{kerber19}, consisting in empirical-point-spread-function (ePSF) fitting to obtain positions and magnitudes. The main differences with respect to Kerber et al. are: (i) we implemented a second-step photometry stage adapting for the GSAOI data the FORTRAN software KS2 \citep[][and Anderson et al., in preparation]{nardiello18, libralato22}.
KS2 is specifically designed to improve the finding of faint sources (by using all images at once) and the photometry in crowded environments (by ePSF subtracting all neighbors to each source prior to measure its position and flux); (ii) we used the \emph{Gaia} DR2 catalog \citep{gaia18} to set the reference frame system with x and y axes pointing towards North and West, respectively, and an arbitrary pixel scale of 40 mas pixel$^{-1}$. The photometry was calibrated on the Two Micron All Sky Survey 
\citep[2MASS;][]{skrutskie06} photometric system with a simple zero-point, and corrected for the effects of differential reddening as in \citet{milone12} and \citet{bellini17}. The HST data reduction was performed following  \citet{nardiello18} and \citet{libralato22}, again by means of first- and second-step photometric methods.

Proper motions were obtained following  \citet{libralato21}, summarized as follows. KS2 also delivers single-exposure catalogs, one for each image, containing raw deblended position and flux of each detected source. For each star in each HST and GSAOI KS2-based raw catalog, we used six-parameter local linear transformations to obtain its position onto the reference frame system of the KS2-based HST master catalog. We used only cluster stars to compute the coefficients of these transformations, meaning that our proper motions are measured relative to the bulk motion of the cluster. Also, positions were corrected by applying local adjustments in order to mitigate the possible presence of small, uncorrected systematic residuals \citep[the so-called boresight correction;][]{anderson10}. The final position of a star in each epoch was estimated as the robust average of the transformed positions in all images of that epoch. At odds with \citet{libralato21}, we used the HST and not the \emph{Gaia} catalog as a reference because of the few sources in common between the latter and the GSAOI catalogs that would have prevented us from using a local-transformation approach. This process might seem redundant for the HST data since the KS2-based positions are computed by averaging the KS2 single-exposure raw catalogs once on to the same reference system. However, this is needed because KS2 does not give output positional uncertainties, which are needed to estimate the proper-motion errors and discern between well- and poorly-measured objects. Finally, proper motions were then defined as the displacements multiplied by the pixel scale (40 mas pixel$^{-1}$) and divided by the temporal baseline ($\sim$16.5 yr). Proper-motion errors were computed as the sum in quadrature of the positional errors of the two epochs, again multiplied by the pixel scale and divided by the temporal baseline.

\subsection{Membership probability}\label{sec:membership}

\begin{figure*}[!htb]
\centering
\includegraphics[width=0.99\textwidth]{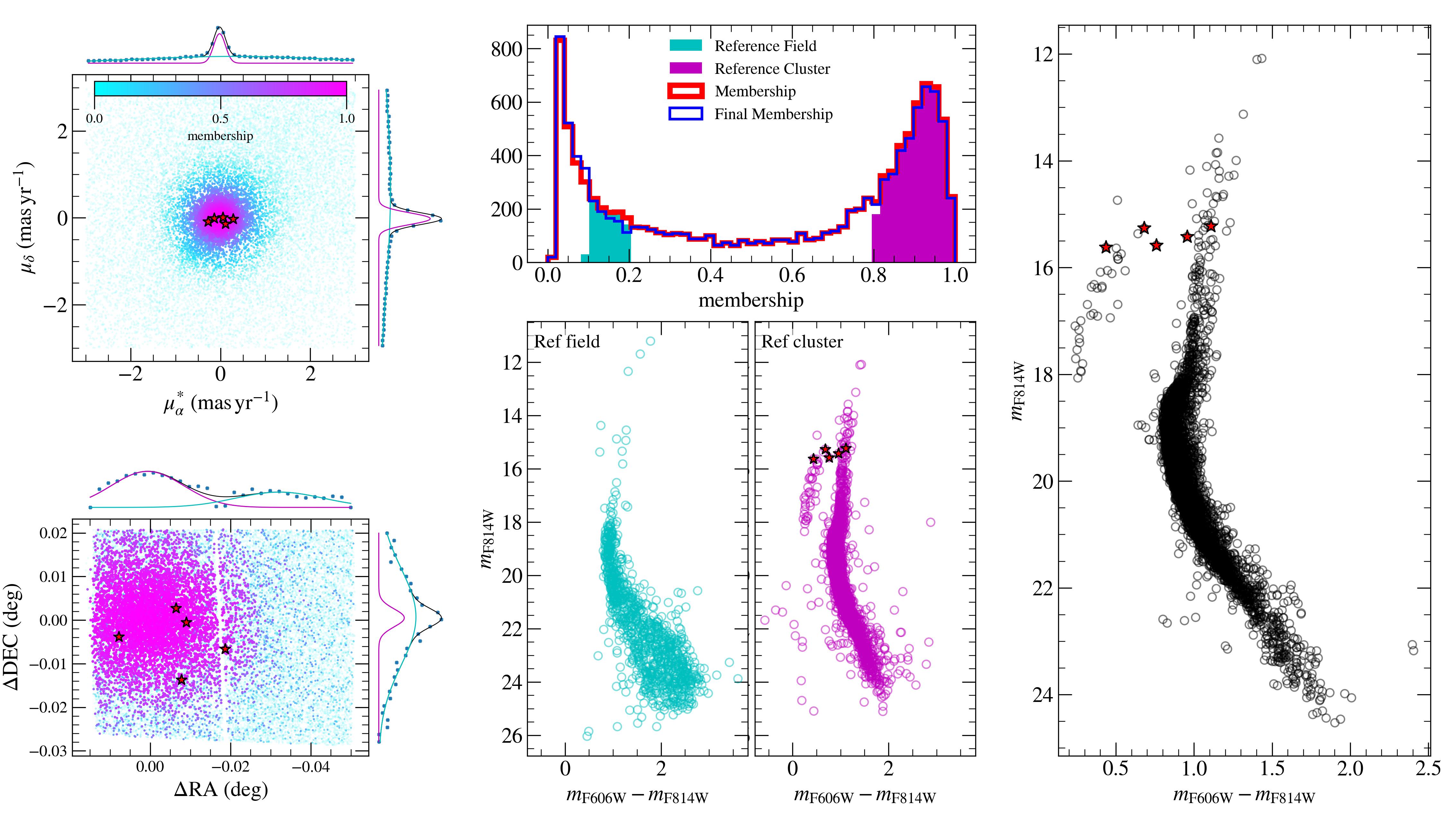}
\caption{{Membership probability derivation. The top left and bottom left panels show VPD and spatial position from the cluster centre coloured by the membership probability value. The corner upper and right panels show the double Gaussian fitting to the distributions. The top middle panel shows the membership distribution before (red) and after (blue) the cleaning method. The cyan region shows the membership range of reference field stars while the magenta one is the cluster members probability region. The bottom middle panels show the reference field and cluster stars on the left and right, respectively. The cleaned CMD is shown in the right panel. The red star symbols show the OGLE RR Lyrae classified as members following the membership method.}}
\label{fig:memb} 
\end{figure*}

{The membership probability for each star was derived using the HST proper motions and RA and DEC values. We fitted a double Gaussian in each dimension of the vector-point diagram (VPD) to find the peak of the cluster distribution and the contribution of the high-spread background. Each star was assigned a Gaussian probability using the cluster mean and standard deviation in the proper motion distribution (top left panel of Figure \ref{fig:memb}). Since some background and foreground stars can have similar proper motions as the cluster and be wrongly assigned as members, we also combined the proper motion membership with a membership probability concerning the distance from the cluster centre (bottom left panel of Figure \ref{fig:memb}). The top middle panel shows the membership probability for the HST catalog. As cluster members, we assume all stars with probabilities above $80\%$. To remove the outlier stars that survived the membership selection, we performed some steps of the method applied by \cite{maia10} and briefly described as follows (bottom middle panel of Figure \ref{fig:memb}): in the same sky region, a sample of field and cluster stars are selected, here we adopted stars with probabilities between $10\%$ and $20\%$ as the field reference, and probabilities above $80\%$ as cluster members; the CMD is divided in a grid with $\Delta=0.08$ and $\Delta=0.25$ mag in colour and magnitude, respectively, which are equivalent to $3$ times the standard deviation in colour and $6$ times the standard deviation in magnitude; the density of field and cluster points in each cell are then compared; a probability is assigned for each cell as $P = (N_{\rm cluster}-N_{\rm field}) / N_{\rm cluster}$ being $0$ when the number of field stars is larger than the cluster stars and removing cells with only one cluster star. In our case, this method was applied to remove as many outlier stars as possible. Therefore, the reference field must not be highly populated. The final CMD in the right panel of Figure \ref{fig:memb} appears less contaminated after the cleaning method. The combined GSAOI+HST catalog is then obtained by cross matching both catalogs.}

\subsection{Differential reddening correction}

{We corrected the data from differential reddening using the method commonly applied in the literature \citep{piotto99,alonsogarcia12,milone12,hendricks12,bellini17} and briefly described as follows: a sample of reference stars is selected, sometimes between the magnitudes where the angle between the reddening vector and the CMD is higher; a fiducial line is derived from this selection of stars; the distance of each reference star and the fiducial line is measured along the reddening vector; this distance is translated into the reddening of each reference star; the final step is, for each star in the catalogue, a number of neighbour stars of the reference sample are selected and the final reddening of the specific star is the median of the reddening distribution of the neighbours. Some authors perform the method several times until the variation on the fiducial line becomes negligible. The differential reddening correction aims to reduce the spread along the CMD, bringing the stars as close as possible to the fiducial line. Therefore, creating the fiducial line is a crucial step for the differential reddening correction.}

\begin{figure*}
\centering
\includegraphics[width=0.99\textwidth]{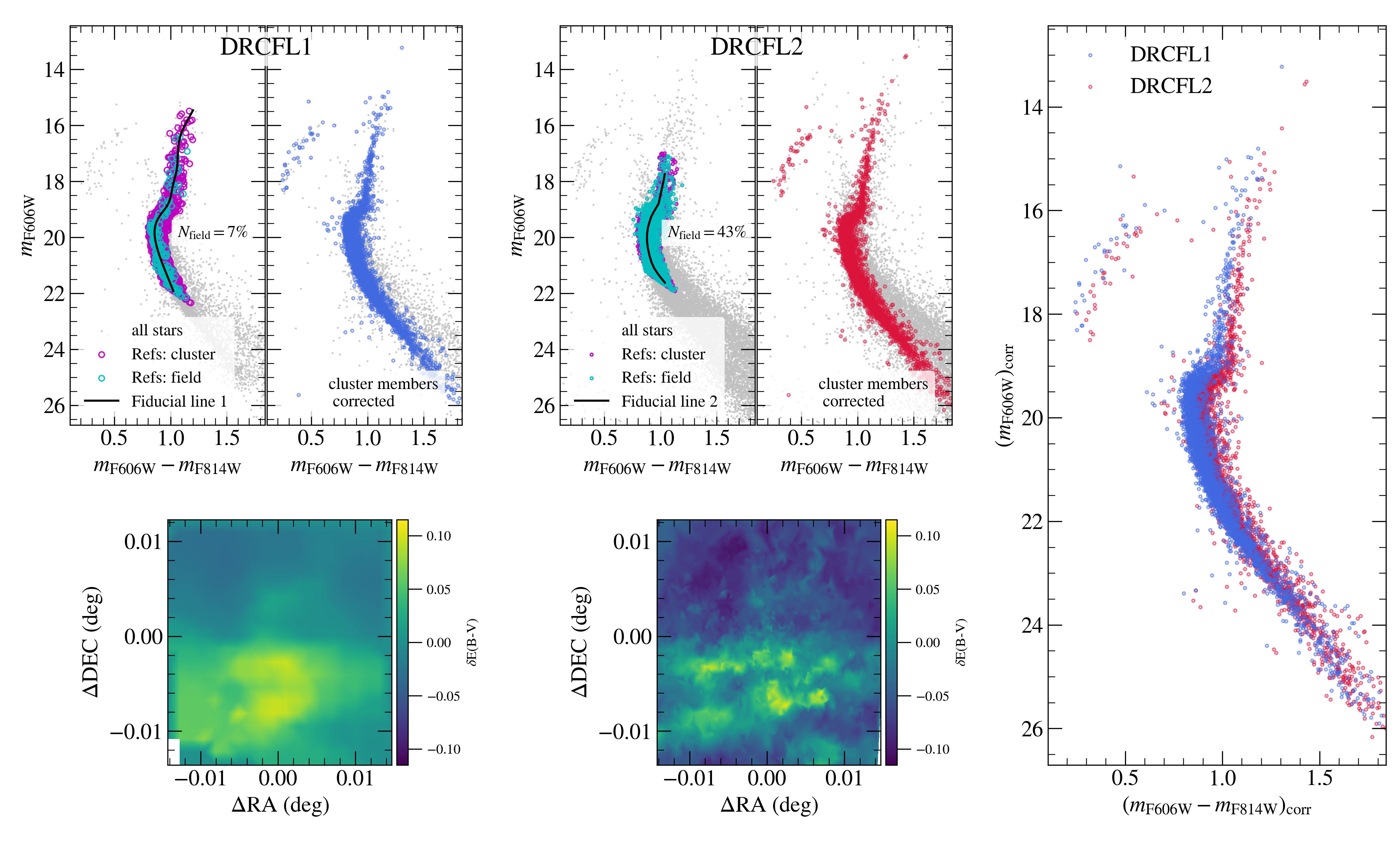}
\caption{{Differential reddening correction procedure. The top left panels show the CMDs with the selected reference stars (field as cyan dots and cluster as magenta dots, same as Figure \ref{fig:memb}) to generate the fiducial line (black line, left) and the corrected CMD (right). The differential reddening maps are at the bottom with the positions with respect to the cluster centre. Both density maps have the same colour code and scale and are limited to the GSAOI FoV. On the left panel is shown the differential reddening correction using the fiducial line constructed from the sample within the GSAOI FoV (DRCFL1), and in the middle panel the differential reddening correction constructing the fiducial line using stars within the HST/ACS FoV. The right panel directly compares the cluster member stars using DRCFL1 (blue) and DRCFL2 (red).}}
\label{diff} 
\end{figure*}

We derived the fiducial line in two different situations to show its influence on the final result of the differential reddening correction (Figure \ref{diff}). First, we selected reference stars in the final combined GSAOI+ACS catalog, which is limited to the GSAOI FoV of $85\times85$ arcsec$^2$ (left panels of Figure \ref{diff}). In this representation, cyan dots denote field stars, while magenta dots represent cluster stars, consistent with Figure \ref{fig:memb}. Since the half-light radius of the cluster is $\sim 45.1$ arcsec \citep[][]{baumgardt20}, the GSAOI FoV has more cluster members as field stars, which is reflected in our selection of reference stars where we have $7\%$ of field stars (left panels of Figure \ref{diff}). The differential reddening map obtained from this less contaminated selection is shown in the bottom-left panel of Figure \ref{diff}. Along this work, we will call this differential reddening correction from fiducial line 1 as DRCFL1.

In another test, we selected the reference stars on the entire ACS/WFC FoV (middle panels of Figure \ref{diff}). Since the centre of the cluster is imaged in one of the two ACS/WFC chips (see Figure \ref{fig:hst_image}), the ACS/WFC images cover a large portion of the field outside the half-light radius of the cluster and, thus, more field stars are present in our catalog. Indeed, for this reference sample, we have $43\%$ of field stars. We corrected for the effects of the differential reddening as before and obtained the map shown in the bottom-middle panel of Figure \ref{diff}. For a direct comparison with the previous test, DRCFL1, the panel only shows the same region covered by the GSAOI observations. We will refer to this differential reddening correction from fiducial line 2 as DRCFL2.

\begin{figure}[htb!]
\centering
\includegraphics[width=0.99\columnwidth]{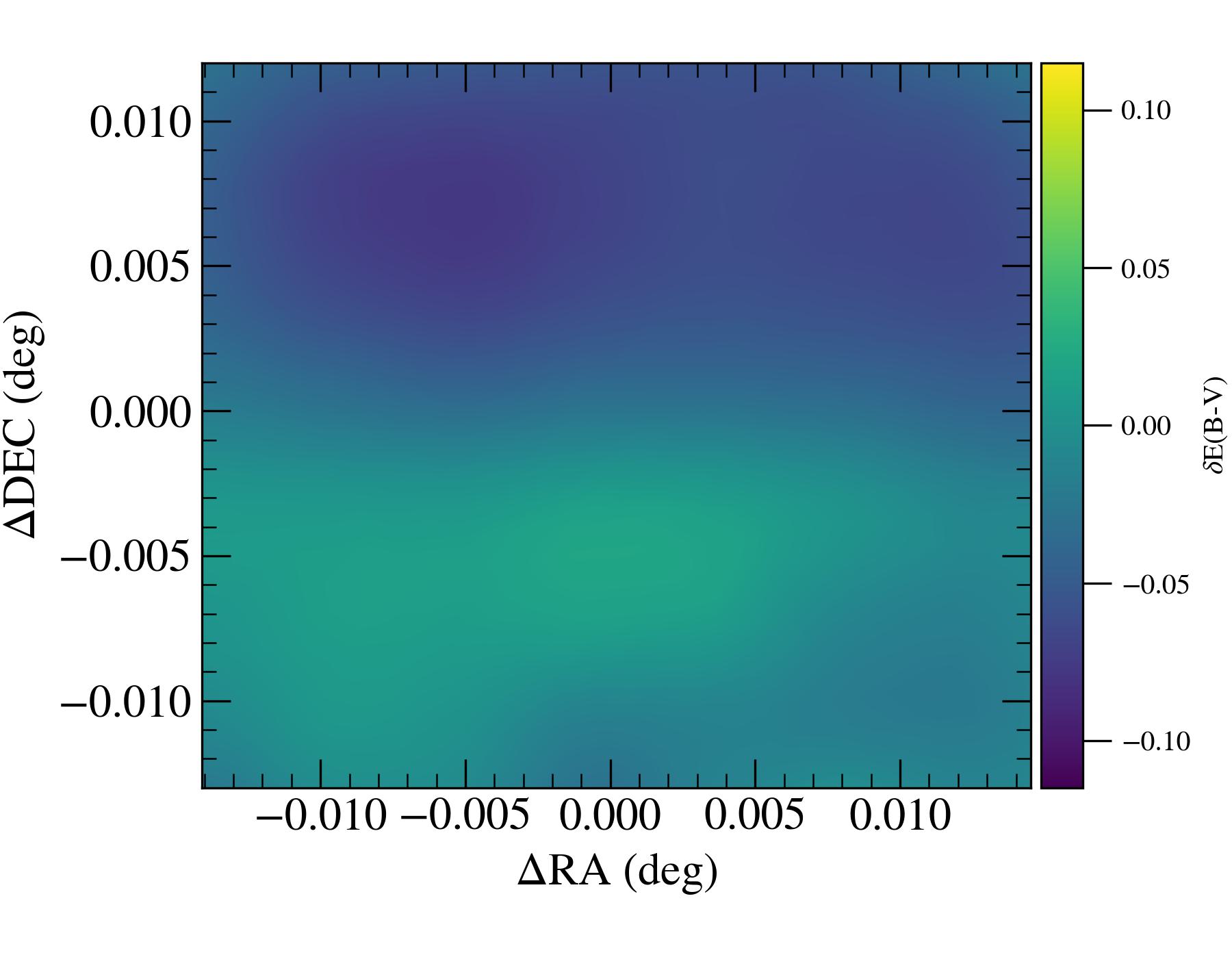}
\caption{{Differential reddening map for NGC~6558 by \cite{alonsogarcia12}. The density map has the same colour code and scale, and region as Figure \ref{diff}.}}
\label{fig:ag_map} 
\end{figure}

A similar approach as DRCFL2 was used by \cite{alonsogarcia12} for NGC~6558, from observations in $B$, $V$, and $I$ bands obtained with the \emph{Magellan} $6.5$ m Telescope and the HST. They covered a circular area of radius approximately $217$ arcsec from the centre of the cluster, covering almost the same area as ACS/WFC FoV. Their resulting differential reddening map is shown in Figure \ref{fig:ag_map}. For a direct comparison, the three differential reddening maps cover the same area (GSAOI FoV) and have the same colour scale.

Both DRCFL1 and DRCFL2 maps show a consistent general behaviour as observed in \cite{alonsogarcia12}: the southern region of the cluster exhibits positive differential reddening, whereas the northern region experiences negative differential reddening. Nevertheless, a systematic variation of approximately $+0.05$ mag is observed in the differential reddening when utilizing GSAOI FoV stars (DRCFL1). This discrepancy arises from a $0.05$ mag difference in the turn-off (TO) colour between the two fiducial lines, where the TO of fiducial line 2 is 0.05 magnitudes redder than the TO of fiducial line 2. The direct comparison of the resulting CMD using DRCFL1 (blue) and DRCFL2 (red) is shown in the right panel of Figure \ref{diff}. This visualization underscores the pivotal role of the fiducial line, particularly for clusters located within the galactic disc or projected towards the Galactic bulge, where the density of background stars is significantly higher, making the fiducial line redder.

\begin{figure*}[htb!]
\centering
\includegraphics[width=0.99\textwidth]{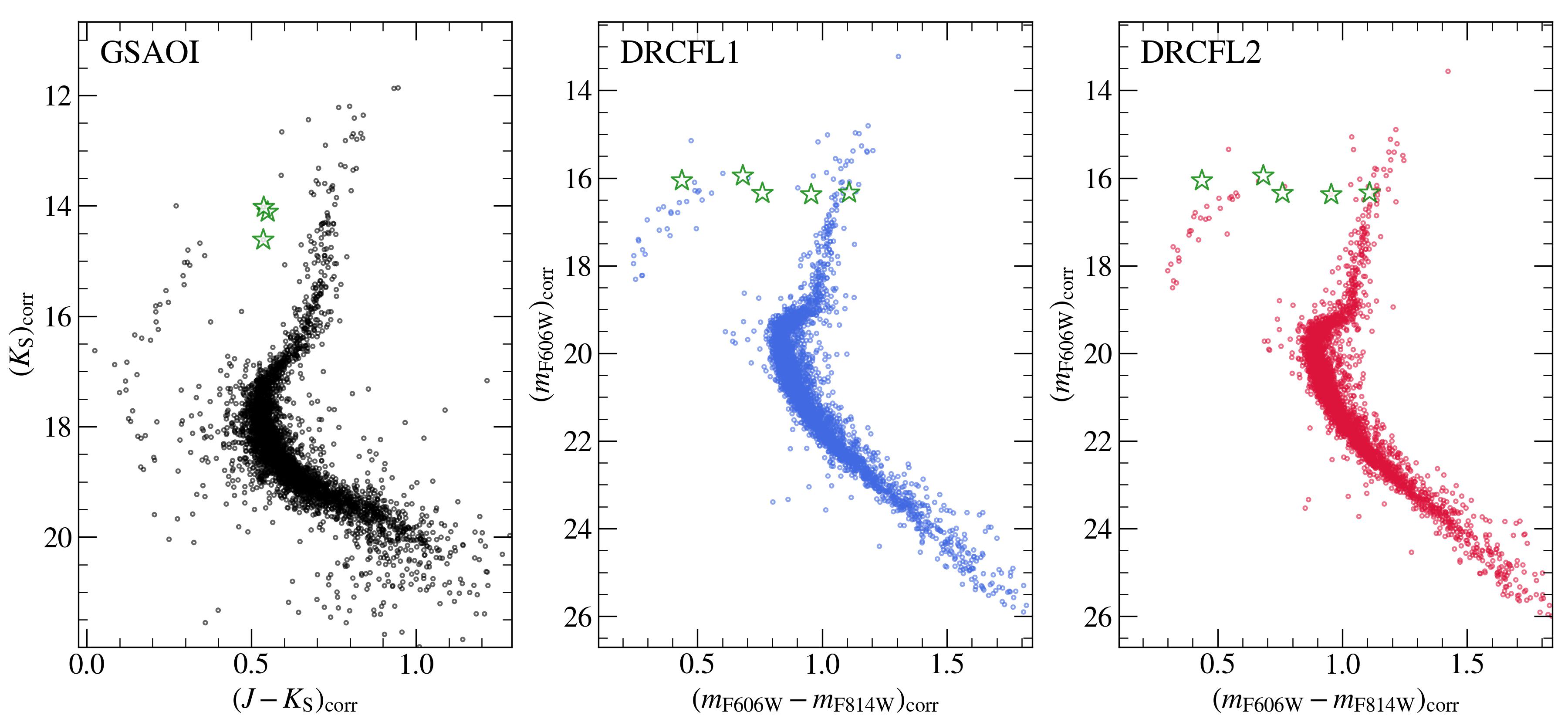}
\caption{{Final CMDs with selected cluster members. The left panel is the GSAOI [$K_S$,$J-K_S$] CMD. The middle and right panels show the HST/ACS [$m_{\rm F606W}$,$m_{\rm F606W}-m_{\rm F814W}$] employing the DRCFL1 and DRCFL2, respectively. The green star symbols are the RR Lyrae members of the cluster within the GSAOI FoV.}}
\label{fig:cmds} 
\end{figure*}

For the subsequent analysis, we adopt the NGC~6558 results obtained by employing the DRCFL1, which takes into account less field star contamination in the differential reddening correction procedure. Nevertheless, we will compare them with the results using the DRCFL2 to emphasize the possible biases resulting from the contamination by background field stars on the fiducial line creation. Figure \ref{fig:cmds} shows the three CMDs that will be employed in this work: GSAOI [$J$, $J-K_S$] (left panel); HST/ACS [$m_{\rm F606W}$, $m_{\rm F606W}-m_{\rm F814W}$] using DRCFL1 (middle panel) and using DRCFL2 (right panel).

\section{RR Lyrae magnitude level}\label{sec:rrlyr}

RR Lyrae stars from \cite{clement01} and OGLE-IV \citep{soszynski14} databases were used as in \citet{oliveira22}, in order to provide a prior distribution to the cluster distance.
\citet[updated in Apr 2016]{clement01} contains 11 RR Lyrae with periods, amplitudes, and mean magnitudes in the $I$ band, with a note that all the data comes from OGLE-IV in this case.
\citet{soszynski14} lists seven RR Lyrae within a radius of $2.1^\prime$, but a wider search of $10^\prime$ in OGLE Collection of Variable Stars\footnote{\url{http://ogledb.astrouw.edu.pl/~ogle/OCVS/}} returns 22 RR Lyrae, with the 11 more central ones in common with \citet{clement01}.
The latter provided time-series photometry and mean $I$ magnitudes from $\sim 660$ epochs, and additional mean $V$ magnitudes from 42 epochs.

\begin{figure}[htb!]
\centering
\includegraphics[width=0.98\columnwidth]{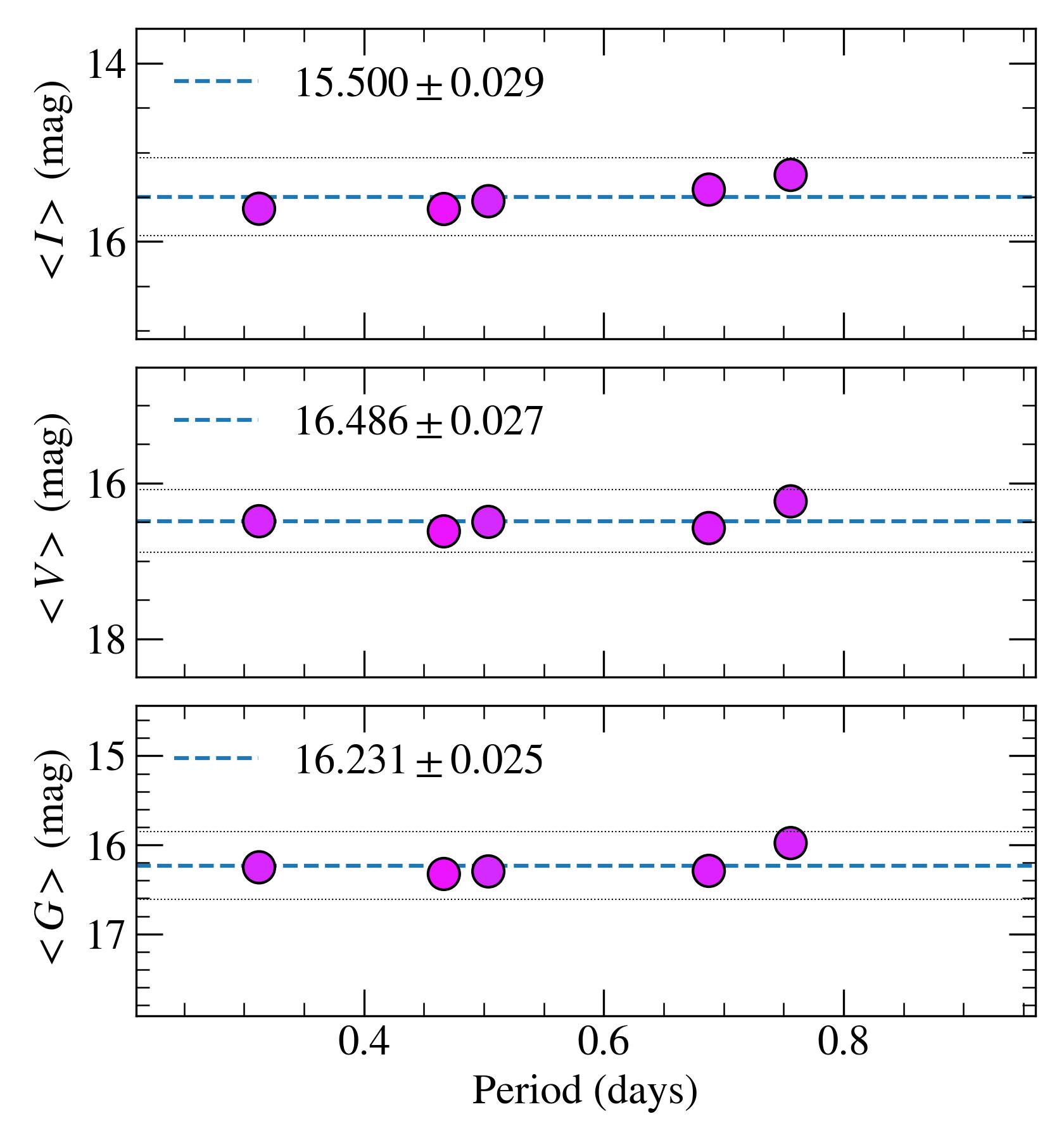}
\caption{Mean $I$, $V$, and $G$ magnitudes versus the period of the RR Lyrae sample. The dashed lines mark the weighted average given in Table~\ref{tab:rrlyr} and the dotted lines show the standard error.}
\label{pm-rrlyr} 
\end{figure}

{A cross-match with our HST/ACS catalog retrieved 5 of the 22 RR Lyrae with a membership probability (derived in Section \ref{sec:membership}) larger than $80\%$, of which three are within the GSAOI FoV.
A weighted average of the mean RR Lyrae magnitudes was obtained with membership values as weights, as presented in Table~\ref{tab:rrlyr} for the different filters. Figure~\ref{pm-rrlyr} presents the mean $V$ and $I$ magnitudes as a function of the period.}

\begin{table}[!ht]
    \centering
    \caption[4]{Parameters of 5 RR Lyrae classified as members of NGC~6558 from their membership probabilities. The last column is the weighted average of the RR Lyrae mean magnitudes in different photometric systems. }
    \resizebox{\columnwidth}{!}{%
    \begin{tabular}{r|ccccc|c}
    \hline
    \noalign{\smallskip}
        \hbox{OGLE ID}           &   \hbox{14866} &  \hbox{14886} &   \hbox{14888} &   \hbox{14892} &   \hbox{14912} & $<\,\,>$ \\ 
        \noalign{\smallskip}
        \hline
        \noalign{\smallskip}
        \hbox{RA}                       & $272.555$ & $272.565$ & $272.566$ & $272.568$ & $272.582$ & --- \\ 
        \hbox{DEC}                      & $-31.771$ & $-31.765$ & $-31.778$ & $-31.762$ & $-31.768$ & --- \\ 
        \hbox{$\mu_{\alpha}^{*}$}       & $ -0.138$ & $  0.113$ & $  0.055$ & $  0.281$ & $ -0.273$ & --- \\ 
        \hbox{$\mu_{\delta}$}           & $ -0.009$ & $ -0.139$ & $  0.008$ & $ -0.026$ & $ -0.087$ & --- \\ 
        \hbox{$I$ (OGLE)}               & $ 15.548$ & $ 15.636$ & $ 15.633$ & $ 15.253$ & $ 15.418$ & $15.50\pm 0.03$\\ 
        \hbox{$V$ (OGLE)}               & $ 16.498$ & $ 16.619$ & $ 16.489$ & $ 16.234$ & $ 16.577$ & $16.49\pm 0.03$\\ 
        \hbox{$G$ (\emph{Gaia})}        & $ 16.297$ & $ 16.326$ & $ 16.251$ & $ 15.979$ & $ 16.291$ & $16.23\pm 0.03$\\ 
        \hbox{$m_{\rm F606W}$}          & $ 16.058$ & $ 16.331$ & $ 16.343$ & $ 15.942$ & $ 16.377$ & $16.21\pm 0.04$\\ 
        \hbox{$m_{\rm F814W}$}          & $ 15.622$ & $ 15.224$ & $ 15.583$ & $ 15.261$ & $ 15.422$ & $15.42\pm 0.03$\\ 
        \hbox{$J$}                      & ---       & $ 15.153$ & ---       & $ 14.560$ & $ 14.660$ & $14.79\pm 0.08$\\ 
        \hbox{$K_S$}                    & ---       & $ 14.616$ & ---       & $ 14.024$ & $ 14.111$ & $14.25\pm 0.08$\\ 
        \hbox{Period}                   & $  0.504$ & $  0.467$ & $  0.312$ & $  0.756$ & $  0.687$ & --- \\
        \hbox{Type}                     & RRab      & RRab      & RRc       & RRab      & RRab      & --- \\
        \hbox{$\mathcal{P_{\rm memb}}$} & $  0.833$ & $  0.923$ & $  0.846$ & $  0.845$ & $  0.865$ & --- \\
    \noalign{\hrule\vskip 0.1cm} 
    \label{tab:rrlyr}
    \end{tabular}}
\end{table}

{The derived magnitudes are converted to distance modulus in each iteration of the isochrone fitting using the $M_I-$ and $M_V-\rm{[Fe/H]}$ relations from \citet{oliveira22} obtained from BaSTI $\alpha$-enhanced models \citep{pietrinferni21} for the zero-age horizontal branch, providing a variable prior in distance which vary with the reddening and metallicity of each Markov chain Monte Carlo (McMC) iteration. The use of the magnitude level appears to be a better constraint to the distance because the assumed metallicity, reddening, and $R_V$ play a key role in the final distance. For example, assuming a metallicity of ${\rm [Fe/H]} = -1.17 \pm 0.10$ \citep{barbuy18b} the calibrations provide apparent distance moduli of $(m-M)_I= 15.30 \pm 0.10$ and $(m-M)_V = 15.86 \pm 0.11$. Converting them to absolute distance moduli using $E(B-V)=0.38 \pm 0.05$\,mag from \citet{barbuy18b} the distances are $8.47 \pm 0.52$ and $8.63 \pm 0.73$\,kpc are obtained respectively from $I$ and $V$, whereas distances smaller by $0.4-0.7$\,kpc are derived with $E(B-V)=0.44$\,mag from \citet[2010 edition]{harris96}. All values are obtained assuming R$_V=3.1$. Therefore, keeping the prior on the magnitude value of the RR Lyrae, we avoid an erroneous distance determination to the detriment of some specific set of metallicity, reddening, and $R_V$.}

{\cite{oosterhoff39,oosterhoff44} classified the MW GCs into two classes based on the mean of the pulsation periods of their RR Lyrae stars: Oosterhoff type I (OoI) and Oosterhoff type II (OoII) group. In a nutshell \citep[see also][]{prudil19a,prudil19b,martinez-vazquez21}, OoI GCs have metallicity values above $-1.7$ and contain mainly RRab stars with a short average pulsation period, commonly around $0.55$ day. The OoI GCs also present a low ($\sim20\%$) of RRc type RR Lyrae. In turn, OoII GCs contain a significant number of RRc stars ($\sim 40\%$), are more metal-poor than $-1.7$, and their RRab RR Lyrae sample shows an average pulsation period of $\sim0.65$ day. The mean period of RRab stars of NGC~6558 is $0.61\pm0.04$ days (Table \ref{tab:rrlyr}). Even though the number of RR Lyrae stars selected as members of the cluster is low, we can estimate that the sample has $20\%$ of first-overtone pulsators (RRc). The average period of RRab and number of RRc in our sample together with the cluster metallicity of $-1.18$ could indicate that NGC~6558 is a OoI GC. Nevertheless, taking into account the low statistics, the average period of RRab stars could also point out to OoII type if the number of RRc stars  would increase with more observations. }

%
%
%
%

\section{Isochrone fitting}

The fundamental parameters age, metallicity, distance, reddening, total-to-selective extinction ratio R$_V$, and binary fraction were derived using the SIRIUS code \citep{souza20}. The code employs the Bayesian method of McMC to obtain probability distributions for each parameter. The code compares the observed CMD with synthetic CMDs constructed from each set of parameters randomly drawn during the fitting process.

To construct the synthetic diagrams, SIRIUS utilizes isochrones from the Dartmouth Stellar Evolution Database \citep[DSED; ][]{dotter08}, which initially spans ages between 10 Gyr and 15 Gyr with intervals of 0.1 Gyr, and metallicities between $-2.30$ and $0.0$ with intervals of $0.01$ dex. However, the code interpolates different isochrones during the fitting process to obtain the model with exact values. With the model constructed, the code adopts the initial mass function of \cite{chabrier03} to draw \emph{N} mass values. These values are then interpolated to obtain the corresponding magnitude values.

A fraction of these \emph{N} mass values are associated with spatially unresolved photometric binary stars (stars close enough to have their flux overestimated). While the first star has mass m$_A$, the hypothetical secondary star will have mass m$_B$ calculated as m$_B = q \cdot m_A$, where q is the mass ratio randomly selected between $0$ and $1$. The calculation of the final magnitude of binary stars is computed from the sum of the flux of each star:

\begin{equation}
M = -2.5\log\left(10^{(-M_A/2.5)} +10^{(-M_B/2.5)}\right)
\end{equation}

With the magnitudes of each star calculated, an error function is applied to spread the points in the CMD. The error function is derived from the observed data by calculating the median error in magnitude bins. The error function is shifted to the position of the theoretical isochrone from the turn-off point to ensure no bias in the position of the synthetic diagram relative to the observed one. Thus, whatever set of parameters is drawn, the error function will generate a synthetic CMD.

After spreading the stars according to the error function, the code calculates the extinction coefficients following the extinction law (R$_V$) of each iteration. The extinction law is interpolated from the curves of \citet{cardelli89}, and the extinction coefficients for the $J$, $K_S$, F606W, and F814W bands are obtained from this. Then, each absolute magnitude is converted to apparent magnitude as follows:

\begin{equation}
m_{\lambda} = M_{\lambda} + (m-M)_0 + R_V \cdot E(B-V) \cdot  A_{\lambda} 
\end{equation}

where (m-M)$_0$ is the intrinsic distance modulus, A$_{\lambda}$ is the extinction coefficient in band $\lambda$. 

With the synthetic diagram already with the apparent magnitudes, a luminosity function is applied to reproduce the observation conditions to which the data were subjected. The luminosity function is calculated as the number of stars in each magnitude bin. When applied to the synthetic diagram, some stars will be excluded to make the synthetic diagram more similar to the observed one in terms of the number of stars at each magnitude.

When the R$_V$ is a free parameter, the isochrone fitting is performed simultaneously for the [$J$, $J-K_S$] and [$m_{\rm F606W}$, $m_{\rm F606W}-m_{\rm F814W}$] diagrams. Simultaneous fitting allows the determination of the extinction law since it will force the adjustment of a single set of distance and reddening for the two diagrams. {Another advantage of this approach is that the age determination is not biased by the choice of the photometric system.}

As a proxy of distance, we are employing a prior to the distance according to the magnitude level of RR Lyrae stars. For each McMC interaction, the values of metallicity and reddening are used for the M$_I - $[Fe/H] calibration by \citet{oliveira22}.

\begin{equation}
    M_I = A +B\cdot{\rm [Fe/H]} + C\cdot{\rm [Fe/H]}^2.
\end{equation}
where $A=(0.619\pm0.028)$, $B=(0.455\pm0.063)$, and $C=(0.075\pm0.030)$. Then, the expected distance for the previously RR Lyrae magnitude level, reddening, and metallicity is calculated for each interaction. This calculated distance is then compared with the distance value for the specific McMC interaction.

The McMC was applied using the \texttt{Python} library \texttt{emcee} \citep{emcee} and the \texttt{PyDE}\footnote{\url{https://github.com/hpparvi/PyDE}}, a global optimization that uses differential evolution. The likelihood function is an adaptation of \cite{tremmel13}, which is basically a Poisson distribution comparing the number of observed and synthetic stars in a colour and magnitude bin. The adaptation includes a simple isochrone fitting component comparing the isochrone to the two-dimensional distribution of observed points. This change allows the code to reach the stability of the distance and reddening values more quickly. Meanwhile, the likelihood component that compares the synthetic diagram with the observed one allows for better age and binary fraction calculation.

\begin{figure*}[ht!]
    \centering
    \includegraphics[width=0.99\textwidth]{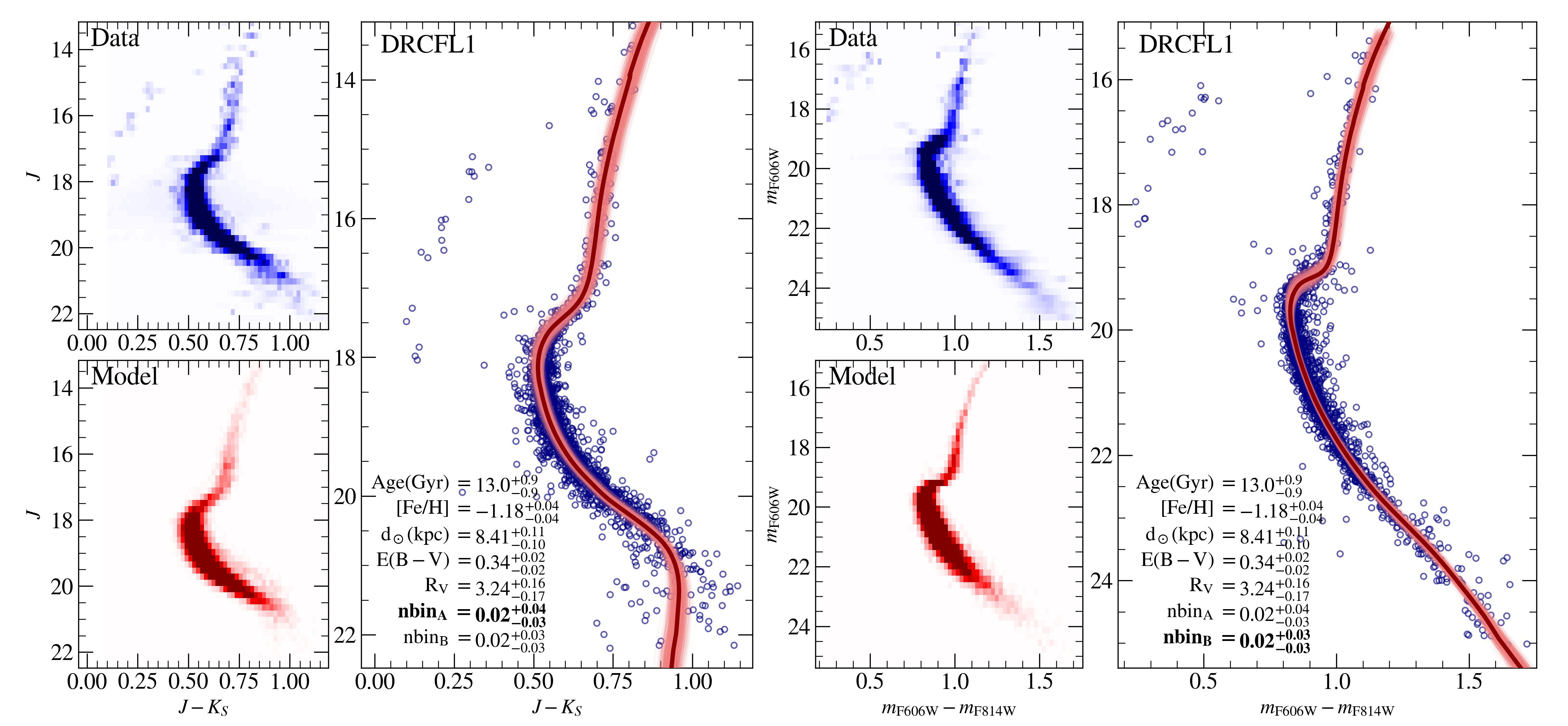}
    \caption{Simultaneous isochrone fitting using the differential reddening correction from fiducial line 1 (DRCFL1). The left panels show the NIR GSAOI CMD, and the right panels show the optical F606W/F814W HST CMD. The small panels show the Hess diagram of the data (upper) and model (best-fit synthetic CMD, lower). In the bigger panel, the blue dots are the cluster member stars, and the red line represents the isochrone of the best synthetic CMD fit and the thin lines are solutions within the errors. The values obtained from the simultaneous fitting are shown in the bottom left corner of the bigger panels, with the respective binary fraction value highlighted.}
    \label{fig:best-fit_FL1}
\end{figure*}

\begin{figure*}
    \centering
    \includegraphics[width=0.99\textwidth]{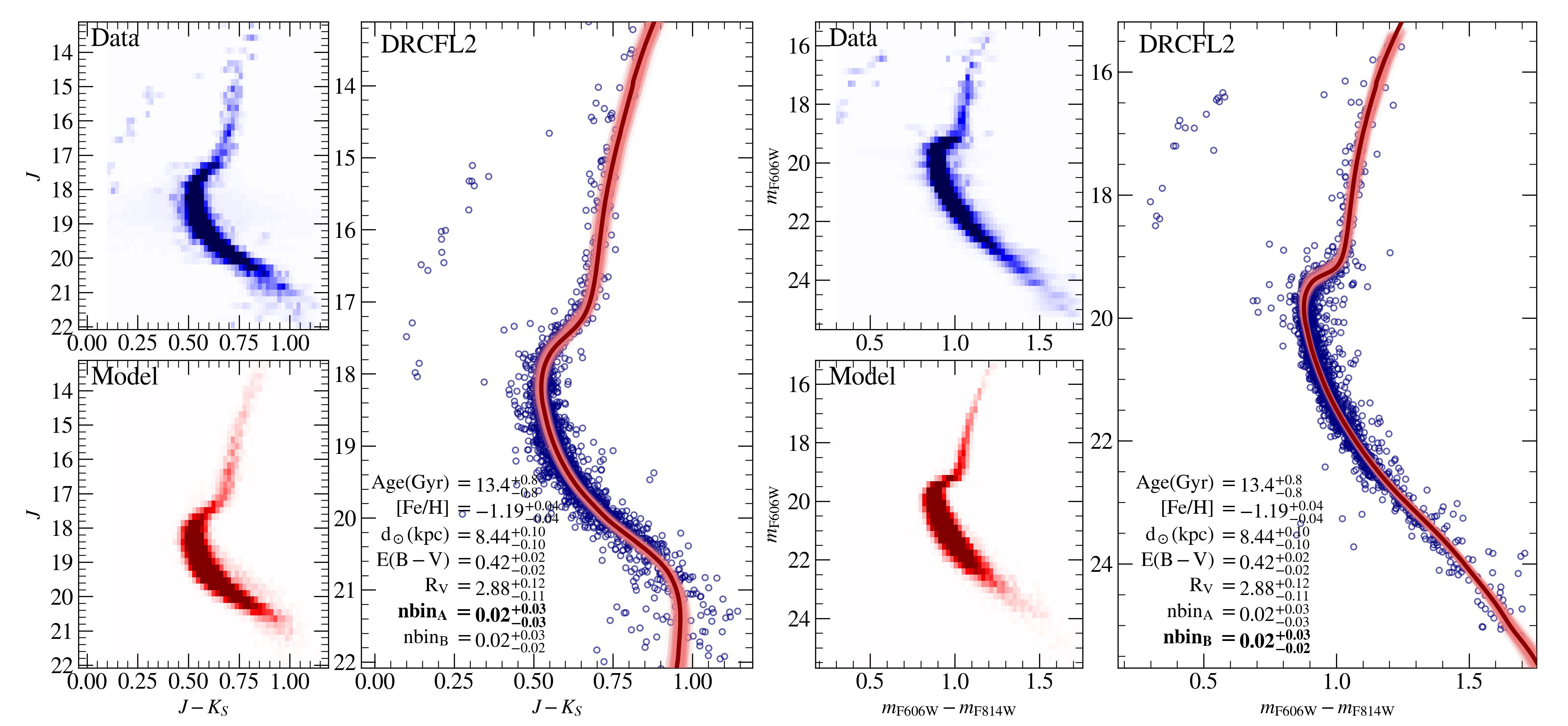}
    \caption{Same as in Figure \ref{fig:best-fit_FL1} but using the differential reddening correction from fiducial line 2 (DRCFL2).}
    \label{fig:best-fit_FL2}
\end{figure*}

In the isochrone fitting procedure below, we applied the metallicity from \cite{barbuy18a} and the RR Lyrae magnitude level from Section \ref{sec:rrlyr} as priors. Figures \ref{fig:best-fit_FL1} and \ref{fig:best-fit_FL2} show the simultaneous isochrone fitting on the [$J$, $J-K_S$] and the [$m_{\rm F606W}$, $m_{\rm F606W}-m_{\rm F814W}$] CMDs (the corner-plots are shown in the Figure \ref{fig:corners}). Figure \ref{fig:best-fit_FL1} shows the HST CMD with the DRCFL1 and Figure \ref{fig:best-fit_FL2} adopting DRCFL2. The distance values, d$_\odot = 8.41\pm0.10$ and $8.44\pm0.10$ kpc, respectively, are in excellent agreement between each other. This agreement is, however, due to compensation in the reddening value E($B-V$) and the extinction law R$_V$ parameter. The E($B-V$) values are $0.34\pm0.02$ and $0.42\pm0.02$ for DRCFL1 and DRCFL2, respectively. More important, is the difference in the R$_V$ values of $3.2$ and $2.9$. The low R$_V$ is expected because the data of DRCFL2 include background Galactic bulge field stars that are affected by dust \citep{minniti14,nataf16}. Removing the Galactic bulge field stars, as is the case of DRCFL1, the relatively high R$_V = 3.2$ is compatible with the Galactic latitude of NGC~6558 ($l = -6.02 ^\circ$), which is outside the dark lane region \citep{minniti14,nataf16}. 

Finally, the difference in the derived ages, $13.0\pm0.9$ and $13.4\pm0.8$ Gyr, reflects the different morphology of the respective fiducial line and the difference in the MSTO colour ($\sim 0.05$ mag).

\section{Orbital parameters}
We performed the orbital integration of NGC~6558 in order to derive the orbital parameters apogalactic distance ($<r_{\rm apo}>$) and perigalactic distance ($<r_{\rm peri}>$), eccentricity ($ecc = (<r_{\rm apo}> - <r_{\rm peri}>)/(<r_{\rm apo}>+<r_{\rm peri}>)$), and the maximum absolute height relative to the disc ($<|Z|_{\rm max}>$). 

For the orbital integration, we adopted the Milky Way model from \cite{portail17} by using the analytical approximation given by \cite{sormani22} with the Action-based GAlaxy Modelling Architecture \citep[AGAMA;][]{vasiliev19}, a Python package for orbital integration\footnote{\url{https://github.com/GalacticDynamics-Oxford/Agama}}. The Galactic model of \cite{portail17} is composed of the x-shape peanut/bulge bar, long bar, central concentration mass, disk, and dark-matter halo. This Galactic model described well the complexity of the inner Galactic region (within $\sim 5$ kpc), although further out is not very realistic. For the case of NGC 6558, this potential is the most suitable since this cluster is supposed to belong to the Galactic bulge \citep{bica16,perez18,perez-villegas20}.

In this model, the position of the Sun is at a distance R$_0=8.2$ kpc \citep{bland-hawthorn16} from the Galactic centre. The bar rotates with a pattern speed of $39$ km s$^{-1}$ kpc$^{-1}$ and is oriented at an angle of $28^o$ with the Sun–Galactic centre line of sight. The mass distribution of the model has a local circular velocity of $V(R_0)=239.2$ km s$^{-1}$. The peculiar motion of the Sun with respect to the local circular orbit is ($U_0$, $V_0$, $W_0$)$=$($11.1$, $12.24$, $7.25$) km s$^{-1}$ \citep{schonrich10}.

\begin{figure}[ht!]
    \centering
    \includegraphics[width=0.99\columnwidth]{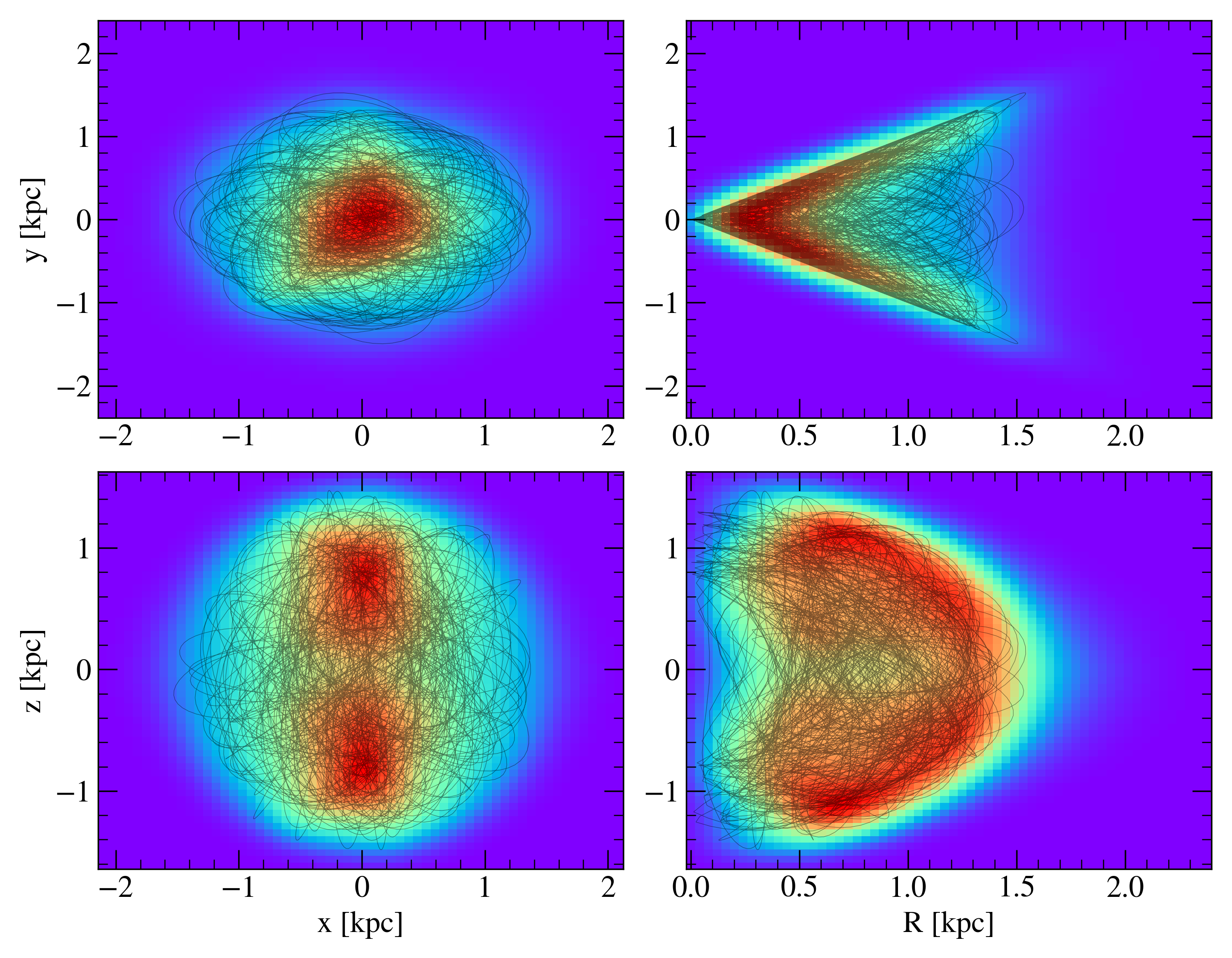}
    \caption{Density probability map for the $x-y$ (upper left), $R-y$ (upper right), $x-z$ (bottom left), and $R-z$ (bottom right) projections of the set of orbits for NGC~6558. Orange corresponds to higher probabilities, and the black lines show the orbits using the main observational parameters.}
    \label{fig:orbits}
\end{figure}

The observational parameters RA, DEC, radial velocity, distance from the Sun, and proper motions in the sky of the cluster, presented in the upper part of Table \ref{tab:orbit},  are converted to Cartesian Galactic phase-space positions and velocities using the \texttt{Astropy} \texttt{Python} package \cite{astropy}. To take into account the uncertainties of the observational parameters, we construct 1000 initial conditions using the Monte Carlo approach, {and integrate the orbits backwards in time for 13 Gyr}. The median orbital parameters and the probabilities of belonging to the Galactic bulge and following the bar orbit are shown in the bottom part of Table \ref{tab:orbit}. From the values of apocentric distance $<r_{\rm apo}>$ and maximum height $<|Z|_{\rm max}>$ together with the classification criteria of \cite{perez-villegas20}, NGC~6558 currently has a $99\%$ probability of belonging to the early Galactic bulge. Our results are in good agreement with the calculations given in \cite{perez-villegas20}. {Figure \ref{fig:orbits} shows the density probability map of the orbits of NGC~6558 in the $x-y$ (upper left), $R-y$ (upper right), $x-z$ (bottom left), and $R-z$ (bottom right) projections. The space region in which the orbits of NGC~6558 cross more frequently are shown in orange, and the black curves are the orbits considering the central values of the observational parameters. Regarding the sense of the orbital motion, we found that NGC 6558 has prograde and retrograde at the same time. The prograde–retrograde orbits, observed also in other MW GCs, are produced by the presence of the Galactic bar and this behavior could be connected with chaos, but still it is not well-understood yet \citep{pichardo04,allen06, perez18}}. 

\begin{table}
    \centering
    \caption{Observational parameters used as input to the orbital integration. Orbital parameters on the present calculations (adopting fiducial lines 1 and 2) compared with results from \cite{perez-villegas20}.}
    \label{tab:orbit}
    \resizebox{\columnwidth}{!}{%
    \begin{tabular}{l|ccc}
        \hline
        \noalign{\smallskip}
         Param & DRCLF1 & DRCLF2 & PV20 \\ 
        \noalign{\smallskip}
        \hline
        \noalign{\smallskip}
         &\multicolumn{3}{c}{Observational Parameters} \\
        \noalign{\smallskip}
        \hline
        \noalign{\smallskip}

        RA (deg)                              & \multicolumn{3}{c}{$272.57397$}   \\
        DEC (deg)                             & \multicolumn{3}{c}{$-31.76451$}   \\
        $V_{\rm rad}$ (km s$^{-1}$)           & \multicolumn{3}{c}{$-194.45\pm2.10$} \\
        $<\mu_{\alpha}^{*}>$ (mas yr$^{-1}$)  & \multicolumn{2}{c}{$-1.75\pm0.02$} & $-1.78\pm0.05$ \\
        $<\mu_{\delta}>$ (mas yr$^{-1}$)      & \multicolumn{2}{c}{$-4.17\pm0.02$} & $-4.12\pm0.04$ \\ 
        d$_{\odot}$ (kpc) & $8.41\pm0.10$ & $8.44\pm0.10$ & $8.26\pm0.53$ \\
        
        \noalign{\smallskip}
        \hline
        \noalign{\smallskip}
         &\multicolumn{3}{c}{Orbital Parameters} \\
        \noalign{\smallskip}
        \hline
        \noalign{\smallskip}

        $<r_{\rm peri}>$ (kpc)           & $0.05\pm0.04$ & $0.04\pm0.02$ & $0.10\pm0.07$ \\
        $<r_{\rm apo}>$ (kpc)            & $1.86\pm0.23$ & $1.78\pm0.22$ & $2.20\pm0.31$ \\ 
        $<|Z|_{\rm max}>$ (kpc)          & $1.45\pm0.04$ & $1.52\pm0.04$ & $1.27\pm0.23$ \\
        $ecc$                          & $0.95\pm0.04$ & $0.96\pm0.03$ & $0.92\pm0.05$ \\
        $\mathcal{P}_{\rm bulge}$ (\%) & $99$ & $99$ & $99$  \\
        $\mathcal{P}_{\rm bar}$ (\%)   & $0.0$ & $0.0$ & $19.3$ \\
       \noalign{\hrule\vskip 0.1cm}               
    \end{tabular} }
\end{table}

\section{Discussion}
\label{sec:disc}

{For the following discussion, we adopted the isochrone fitting results of DRCFL1 as the final parameters for NGC~6558. In Table \ref{family}, we tentatively identify a list of Galactic bulge clusters with characteristics similar to NGC~6558, which are probably the oldest objects in the Galaxy. A list of possible members is also given for clusters that still do not have precise age derivation and/or were classified as thick disk members by PV20. It is important to point out that the classification as a bulge or thick disk member depends crucially on the distance adopted, which may be uncertain with the available data. A summary of the literature data for NGC~6558, including reddening values, age, distance, the corresponding method employed, and metallicity, is given in Table \ref{tab:literature}.}

\begin{table*}
\small
\centering
\scalefont{0.8}
\caption[4]{\label{family}
Family of Galactic bulge globular clusters with a metallicity of [Fe/H]$\sim$-1.1,
and very old age.
}
\resizebox{0.99\textwidth}{!}{%
\begin{tabular}{l@{}cccl@{}ccl@{}ccccc}
\hline
\noalign{\smallskip}
\hbox{Cluster} & \hbox{$l$} &  \hbox{$b$} & \hbox{Age}   & \hbox{Ref.} & \hbox{d$_{\odot}$} &  \hbox{[Fe/H]}  &\hbox{Ref.} & \hbox{E(B-V)} & \hbox{HB} & \hbox{M$_V$}  & \hbox{mass}                & orbit \\
\hbox{}        & \hbox{deg} & \hbox{deg}  & \hbox{(Gyr)} &             & \hbox{kpc}         & \hbox{}         & \hbox{}    & \hbox{}       & \hbox{}   & \hbox{}       & \hbox{$10^{5}$M$_{\odot}$} & \hbox{}   \\ 
\noalign{\smallskip}
\hline
\noalign{\smallskip}
NGC~6256              & $347.79$ &  $+3.31$  & $12.9\pm0.1$     & 1 & $6.40\pm0.64$  & $-1.05\pm0.13$  & 10    & $1.10$ & blue         & $-7.15$ & $0.72\pm0.33$  & B \\
HP~1                  & $357.42$ &  $+2.12$  & $12.8\pm0.9$     & 2 & $6.59\pm0.65$  & $-1.01\pm0.10$  & 11,12 & $1.19$ & blue         & $-6.46$ & $1.11\pm0.38$  & B \\
NGC~6401              & $3.45  $ &  $+3.98$  & $13.2\pm1.2$     & 1 & $7.70\pm0.77$  & $-1.00\pm0.03$  & 10,12 & $0.53$ & red          & $-7.90$ & $1.21\pm0.21$  & B \\
Palomar~6             & $2.09  $ &  $+1.78$  & $12.4\pm0.9$     & 3 & $7.67\pm0.19$  & $-1.10\pm0.09$  & 3     & $1.33$ & red          & $-6.79$ & $0.86\pm0.19$  & B \\
ESO456-SC38 (Djorg 2) & $2.76  $ &  $-2.50$  & $12.7\pm0.7$     & 4 & $8.75\pm0.20$  & $-1.07\pm0.09$  & 13    & $0.89$ & blue         & $-7.00$ & $1.34\pm0.24$  & B\\
NGC~6522              & $1.02  $ &  $-3.93$  & $12.8\pm1.0$     & 5 & $7.40\pm0.19$  & $-1.05\pm0.11$  & 14,15 & $0.48$ & blue         & $-7.65$ & $3.92\pm0.54$  & B \\
AL~3/BH~261           & $3.36  $ &  $-5.27$  & $13.4\pm1.2^{*}$ & 6 & $6.50\pm0.65$  & $-1.09\pm0.05$  & 6,12  & $0.36$ & blue         & $-4.19$ & $0.24\pm0.06$  & B\\
NGC~6626              & $7.80  $ &  $-5.58$  & $12.8\pm1.0$     & 5 & $5.34\pm0.17$  & $-1.29\pm0.01$  & 16    & $0.42$ & blue         & $-8.16$ & $3.69\pm0.38$  & B\\
NGC~6642              & $9.81  $ &  $-6.44$  & $12.7\pm1.1$     & 1 & $8.10\pm0.81$  & $-1.11\pm0.04$  & 13    & $0.42$ & intermediate & $-6.66$ & $0.87\pm0.55$  & B\\
NGC~6717              & $12.88 $ &  $-10.90$ & $13.5\pm0.8$     & 7 & $7.14\pm0.10$  & $-1.26\pm0.07$  & 17    & $0.23$ & blue         & $-5.66$ & $0.26\pm0.02$  & B \\
\noalign{\smallskip}
\hline
\noalign{\smallskip}
\multicolumn{13}{c}{  Possible members} \\
\noalign{\smallskip}
\hline
\noalign{\smallskip}
Terzan~1     & $357.57$ & $+1.00$  & $12.6\pm1.3^{*}$    & 21  &  $6.70\pm0.7$    & $-1.06\pm0.13$  & 10 & $2.48$ & red           & $-4.41$ & $2.23\pm0.66$  & B\\
UKS~1        & $5.13$   & $-0.76$  & $13.0\pm1.2$        & 8   &  $11.10\pm1.80$  & $-0.98\pm0.11$  & 18 & $2.62$ & ---           & $-6.91$ & $0.80\pm0.01$  & --- \\
NGC~6355     & $359.58$ & $+5.43$  & $13.2\pm1.0$        & 0   &  $8.54\pm0.19$   & $-1.39\pm0.08$  & 0  & $0.89$ & blue          & $-8.07$ & $0.99\pm0.03$  & B\\
NGC~6540     & $3.29$   & $-3.31$  & $12.6\pm1.3^{*}$    & 21  &  $5.20\pm0.52$   & $-1.06\pm0.06$  & 13 & $0.60$ & ---           & $-6.35$ & $0.56\pm0.04$  & B\\
NGC~6638     & $7.90$   & $-7.15$  & $12.0\pm1.0$        & 9   &  $10.32\pm1.03$  & $-0.99\pm0.07$  & 17 & $0.07$ & intermediate  & $-7.12$ & $1.24\pm0.23$  & TD\\
NGC~6723     & $0.07$   & $-17.23$ & $12.6\pm0.6$        & 7   &  $8.17\pm0.11$   & $-1.01\pm0.05$  & 20 & $0.05$ & intermediate  & $-7.83$ & $1.57\pm0.13$  & TD \\
\noalign{\hrule\vskip 0.1cm}              
\end{tabular}}
\tablebib{
B: Bulge; TD: Thick disc. References: Ages: 0 \citet[][S23]{souza23};1 \citet[][C21]{cohen21}; 2 \citet[][K19]{kerber19}; 3 \citet[][S21]{souza21};
4 \citet[][Or19]{ortolani19}; 5 \citet[][K18]{kerber18}; 6 \citet[][B21b]{barbuy21b};
7 \citet[][Ol20]{oliveira20}; 8 \citet[][FT20]{fernandeztrincado20}; 9 \citet[][M06]{meissner06}; 21 \citet[][K16]{kharchenko16}.
Metallicity: 10 \citet[][V18]{vasquez18}; 11 \citet[][B16]{barbuy16}; 12 \citet[][Ge23]{geisler23};
13 \citet[][Ge21]{geisler21}; 14 \citet[][B21a]{barbuy21a}; 15 \citet[][FT19]{fernandeztrincado19};
16 \citet[][Vi17]{villanova17}; 17 \citet[][C09]{carretta09}; 18 \citet[][FT20]{fernandeztrincado20};
19 \citet[][E19]{ernandes19}; 20 \citet[][FT21]{fernandeztrincado21b}; 21 \citet[][K16]{kharchenko16}.
Distances are from PV20 except for Pal~6 from S21. Absolute magnitudes are from \citet{harris96}.
Masses are from \citet{baumgardt18}\footnote{\url{https://people.smp.uq.edu.au/HolgerBaumgardt/globular/}}. Note: * indicates that the age was derived from the integrated magnitude and the errors are assumed as 10\% of the age value.}
\end{table*}

\begin{table}
\small
\caption[4]{\label{tab:literature}
Literature ages and metallicities for NGC~6558.}
\begin{tabular}{l@{}ccccc}
\hline
\noalign{\smallskip}
\hbox{E(B-V)} & \hbox{Age} & \hbox{d$_{\odot}$} & \hbox{Method} & \hbox{[Fe/H]}  &\hbox{Reference}  \\
\hbox{} &\hbox{(Gyr)} & \hbox{kpc} & \hbox{} & \hbox{} & \hbox{}   \\ 
\noalign{\smallskip}
\hline
\noalign{\smallskip}
0.43 & --- & 6.6 & 1 & --1.65 & H96  \\
0.50 & --- & 6.3 & 2 & -1.2/-1.6 & R98  \\
--- & --- & --- & --- & -1.5 & D04  \\
0.38 & 14 & 6.4 & 2 & -0.97 & B07 \\
0.44 & --- & --- & 2 & -1.21 & C10  \\
0.44 & --- & 7.4 & --- & -1.32 & H10 \\
--- & --- & --- & --- & -1.03 & S12  \\
--- & --- & --- & --- & -1.012 & D15  \\
--- & --- & 7.4 & 3 & --- & R15 \\ 
0.38 & --- & 8.26$\pm$0.53 & -- & -1.17 & B18  \\
--- & --- & 7.474$^{0.294}_{-0.282}$0.7 & 4 & --- & BV21  \\
--- & --- & --- & -- & -1.15 & GD23  \\
\noalign{\hrule\vskip 0.1cm}               
\end{tabular}
\tablebib{
\citet[][H96]{hazen96}; \citet[][R98]{rich98}; 
\citet[][D04]{davidge04};  \citet[][B07]{barbuy07}; \citet[][C10]{chun10}; \citet[][H10]{harris96};
\citet[][S12]{saviane12}; \citet[][D15]{dias15}; \citet[][R15]{rossi15}; \citet[][B18b]{barbuy18b};
\citet[][BV21]{baumgardt21}; \citet[][VB21]{vasiliev19}; \citet[][GD23]{gonzalez23}; 
Methods for age and distance derivation: 1 RR Lyrae; 2 isochrone fitting; 3 isochrone fitting on proper motion cleaned CMDs; 4 \emph{Gaia} proper motions.}
\end{table}

{The metallicities derived from the isochrone fitting are in good agreement with the value from \cite{barbuy18a,gonzalez23} since it was adopted as a prior. The metallicity of NGC~6558 is therefore consolidated from literature studies, as follows. \citet{barbuy07} carried out a detailed abundance analysis of five stars with spectra obtained with the GIRAFFE spectrograph at the Very Large Telescope (VLT) \citep[see also][]{zoccali08}. They obtained a mean metallicity of [Fe/H]$ = -0.97\pm0.15$ for NGC 6558. From CaT triplet lines based on FORS2/VLT spectra, \citet{saviane12} derived [Fe/H]$ = -1.03\pm0.14$. Also  based on FORS2/VLT spectra, \cite{dias15,dias16}
compared the observed spectra to grids of observed and synthetic
spectra and obtained [Fe/H]$= -1.012\pm0.013$. \citet{barbuy18b} analysed 4 stars of NGC~6558 observed with the UVES spectrograph at the VLT. A radial velocity of $194.45$ km s$^{-1}$ and metallicity of [Fe/H]$ = -1.17\pm0.10$ were obtained. \citet{gonzalez23} analysed 4 stars within the APOGEE-2S CNTAC 2 CN2019B program  (P.I: Doug Geisler)
as part of the CAPOS survey \citep{geisler21}. They obtained a radial velocity of $192.63$ km s$^{-1}$ and a metallicity of [Fe/H]$ = -1.15\pm0.10$. Therefore, the Gaussian metallicity prior with mean $-1.17$ and standard deviation of $0.10$ is reasonable to encompass the literature metallicity derivations.}

{The derived heliocentric distance of the cluster is d$_{\odot} = 8.41^{+0.11}_{-0.10}$ kpc is larger than other data given in the literature. \citet{rich98} observed NGC~6558 with the \emph{New Technology
Telescope} (NTT) at the European Southern Observatory using the ESO Multi-Mode Instrument (EMMI) in the imaging/focal reducer mode, reaching a magnitude of V$\sim20.5$ in the CMD. They derived a lower distance of $6.3$ kpc and metallicity between [Fe/H]$ = -1.2$ to $-1.6$. \citet{baumgardt21} also found a lower distance of d$_{\odot}$ = 7.47 kpc, taking the average distance determinations in the literature, which is in agreement with d$_{\odot} = 7.4\pm0.7$ kpc from \citet{rossi15}
who used two epoch NTT data.
A larger distance of d$_{\odot} = 8.26$ kpc close to our determination was given in \citet{barbuy18b} derived from proper motion cleaned NTT photometry, in [$V$, $V-I$]. }

An important contribution of this work is the simultaneous isochrone fitting using both the NIR and optical photometries. This approach allows the determination of the total to selective coefficient R$_V$ \citep[e.g.][]{pallanca21} by imposing one set of fundamental parameters to fit both CMDs. With the DRCFL1, we derived R$_V = 3.2\pm0.2$, and R$_V = 2.8\pm0.1$ using the DRCFL2. The lower value obtained with DRCFL2 reflects the presence of stars in the dust dark lane \citep{minniti14} given that a larger contribution of field stars in the HST field of view (Figure \ref{fig:hst_image}). The variation in the R$_V$ value also implies a reddening E($B-V$) value variation. In Table \ref{tab:literature}, the reddening E($B-V$) literature values range from 0.38 to 0.50. Interestingly, the isochrone fitting gives the values E($B-V$)$ = 0.34\pm0.02$ and $0.42\pm0.02$ for DRCFL1 and DRCFL2, respectively. This variation again shows that the reddening tends to decrease to the value of E($B-V$)$ = 0.35$ when removing the background Galactic bulge field stars behind the dust lane. {The reddening in infra-red bands $E(J-K_S)$ can be obtained from the $E(B-V)$ and $R_V$ provided by our results using the following equation: $E(J-K_S) = E(B-V)\cdot R_V\cdot(R_J-R_{K_S})$. For the $R_V  = 3.24$ (our result), $R_J=0.29186045$ and $R_{K_S}  = 0.11886802$. Therefore, $E(J-K_S)=0.34\cdot3.24\cdot(0.29186045 - 0.11886802) = 0.19$. The extinction in each band is obtained from $A_J  = R_J \cdot E(B-V)\cdot R_V  = 0.29186045\cdot0.34\cdot3.24 = 0.32$.}

The age determination also differs according to the adopted DRCFL1 or DRCFL2, with values of $13.0\pm0.9$ and $13.4\pm0.9$ Gyr, respectively. There is a trend toward older ages with DRCFL2 because of the higher contamination of Galactic bulge field stars. This interesting trend indicates that the Galactic bulge is even older than NGC~6558. \cite{cohen21}, using the method of relative age, derived an age of $12.3\pm1.1$ Gyr for NGC~6558. They used the same observations as those used in this work, which were reduced with a different pipeline. The values derived in this work and by \cite{cohen21} agree within the errors, even though the median values differ by $0.7$ Gyr.

To insert the results of this work in the context of the Galactic bulge formation, we analysed the sample of moderately metal-poor (MMP, [Fe/H]$\sim -1.10$) GCs of the Galactic bulge, which have available age determination in the literature (Table \ref{family}). We selected the orbital parameters of the clusters from \cite{perez-villegas20}, while for NGC~6558, we considered only the results from the present DRCFL1 (Table \ref{tab:orbit}). 

\begin{figure*}[ht!]
    \centering
    \includegraphics[width=0.99\textwidth]{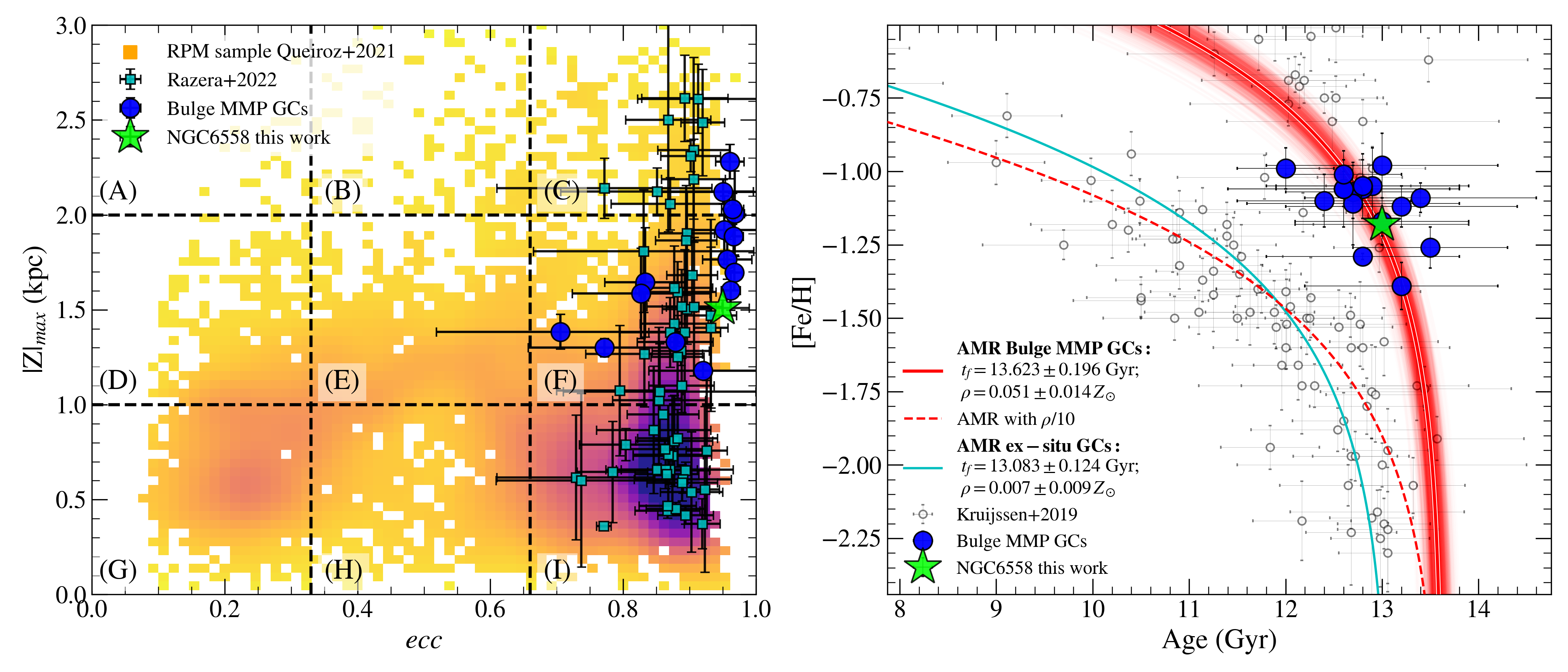}
    \caption{Analysis of the moderately metal-poor side of the Galactic bulge GCs. Left panel: $ecc$ vs. $|Z|_{\rm max}$ plane divided into 9 regions named from A to I. The background density map shows the RPM sample \citep{queiroz21}, the blue dots are the MMP GCs, and the cyan squares are the 58 Galactic bulge spheroid stars from \cite{razera22}. NGC~6558 is the green star symbol. See \cite{perez-villegas20,queiroz21,razera22} for the orbital parameters calculations. Right panel: derivation of the AMR (red line) for the sample of GCs in Table \ref{family} (blue dots), the thin red lines correspond to the AMR varying the mean parameters within errors. The red dashed line shows the AMR assuming a chemical enhancement 10 times slower than the derived for the MMPGCs, while the cyan line is the AMR fitted to the ex-situ GCs. }
    \label{fig:amr}
\end{figure*}

In the left panel of Figure \ref{fig:amr} is the $ecc$ vs. $|Z|_{\rm max}$ plane divided into nine cells named from A to I. This plane is a very good way to split as much as possible the different populations of the inner Galaxy \citep{queiroz21}. The important cells for this work are those with high eccentricity because their shapes are supported by pressure, indicating a spheroidal structure. \cite{razera22} analysed a sample of 58 stars from the reduced-proper-motion (RPM) sample of the inner Galaxy \citep{queiroz21} with metallicity values around the peak of the Galactic bulge metallicity distribution function \citep[$\sim -1.1$;][]{bica16}. These stars are shown in the left panel of Figure \ref{fig:amr} as cyan squares symbols. After a meticulous chemical analysis, the authors concluded that this star sample is a genuine member of a spheroidal structure in the inner Galaxy. The MMP GCs are all located in cells C and F, indicating they are also possible members of the spheroidal Galactic bulge. Moreover, since all MMP GCs are located in cell F, this cell probably represents the oldest population of the inner Galaxy.

In principle, the age-metallicity relation (AMR) of the Galactic bulge MMP GCs shows they are almost coeval among them (right panel of Figure \ref{fig:amr}). The grey dots in the right panel of Figure \ref{fig:amr} compose the sample of 96 Galactic GCs collected by \cite{kruijssen19}. The authors provide an average age for all GCs using the three by \cite{forbes10}, \cite{dotter10,dotter11}, and \cite{vandenberg13}. In order to investigate the old and spheroidal structure of the inner Galaxy (cell F), we fitted the AMR for the MMP GCs using the leaky-box formalism often used in the literature \citep[e.g.][]{massari19,forbes20,limberg22,callingham22} with some modifications:

\begin{equation}
	\displaystyle Z = -\rho \cdot \ln{\left( \frac{t}{t_f} \right)}= Z_\odot \cdot 10^{\rm [M/H]}
\end{equation}
where $t_f$ is the lookback time when the stellar population starts to form stars, $Z_\odot=0.019$ for the solar total metallicity, and $\rho$ is the effective yield of the stellar population. The $\rho$ represents the mass ratio of the new stars formed from the enriched gas expelled from supernovae, being, therefore, a measure of chemical enrichment efficiency of the stellar population and is in units of solar total metallicity ($Z_\odot$). {We are considering the relation ${\rm [M/H]} = {\rm [Fe/H]} + \log_{10}\left( 0.694\cdot10^{\rm [\alpha/Fe]} + 0.306\right)$}. For the MMP GCs of the Galactic bulge we assumed ${\rm [\alpha/Fe]}=+0.4$ \citep{barbuy18a} and ${\rm [\alpha/Fe]}=+0.0$ for the ex-situ GCs \citep{helmi18}. Therefore:
\begin{equation}
  \displaystyle t = t_f \cdot \exp\left( -\frac{Z_\odot}{\rho} \cdot 10^{\rm [Fe/H] + \Delta} \right)
    \begin{cases}
      \Delta = 0.312 & \text{for MMPGCs}\\
      \Delta = 0.0 & \text{for ex-situ GCs}
    \end{cases}       
\end{equation}

{For the MMP GCs Galactic bulge population (Table \ref{family}), we obtained a yield $\rho=0.051\pm0.014\,\,Z_\odot$ and $t_f = 13.623\pm0.196$ Gyr. The derived $t_f$ value indicates that this population is among the oldest ones in the Galaxy since their formation time is close to the age of the Universe of $13.799\pm0.021$ Gyr \citep{planck16}. To investigate the derived effective yield, we also fitted the ex-situ branch of the AMR. The result is the cyan solid line in the right panel of Figure \ref{fig:amr} with $\rho=0.007\pm0.009\,\,Z_\odot$ and $t_f = 13.083\pm0.124$ Gyr. The effective yield is approximately $8$ times smaller than the value derived for the MMP GCs. In the right panel of Figure \ref{fig:amr}, we also present as a red dashed line the AMR with an effective yield 10 times smaller than the MMP GCs and with the same $t_f$. It indicates that the spheroidal structure of the inner Galaxy represented by cell F was formed at the beginning of the Galaxy at $13.623\pm0.196$ Gyr ago and has been chemically enriched approximately ten times faster than the rest of the Galaxy \citep{barbuy18a}.}

\section{Conclusions}

We have used high-quality photometric data from Gemini-South/GSAOI and HST/ACS to derive the parameters distance, reddening, total-to-selective coefficient and age for the Galactic bulge globular cluster NGC~6558. For the isochrone fitting, the metallicity from high-resolution spectroscopy and the apparent magnitude level of RR Lyrae were adopted as priors. 

We demonstrated the importance of the adopted sample of stars in determining the fiducial line for the differential reddening correction. In the NGC~6558 case, when background Galactic bulge field stars contaminate the sample, the isochrone fitting results in a lower R$_V$ and higher extinction, compatible with including Galactic bulge field stars behind a dust lane. Contrarily, removing the Galactic bulge field stars, the results are more compatible with the Galactic latitude of the cluster. The combination of near-infrared and optical data allowed the estimation of the total-to-selective coefficient as R$_{\rm V} = 3.2\pm0.2$ once a fiducial line based on cluster stars with less contribution of field stars was adopted.

The distance of d$_{\odot}$ = 8.41$^{+0.11}_{-0.10}$ kpc obtained is farther than any other in the literature, exemplifying the difficulties for deriving distance values in the literature, even in the \emph{Gaia} era, due to the high reddening in the Galactic bulge \citep[see also][]{vasiliev21}. The reddening of E($B-V$)$ = 0.34\pm0.02$ is the lowest value relative to the literature,
obtained with the differential reddening correction based on reference stars less contaminated by field stars. The age of 13.0$\pm$0.9 Gyr obtained is compatible with the similarly MMP Galactic bulge clusters listed in Table \ref{family}. 

Finally, combining NGC~6558 with the other MMP GCs of the Galactic bulge located within the spheroidal inner structure (cell F in $ecc$ vs. $|Z|_{\rm max}$ plane), we obtained a steep asymptotic trend indicating that the formation time of this population is as late as $13.62$ Gyr and with a chemical enrichment ten times faster than the rest of the MW, making clear that this family of clusters includes the oldest objects in the Galaxy. 

\begin{acknowledgements}
{The authors are grateful to the anonymous referee for his/her suggestions and remarks, which greatly improved the paper.} SOS acknowledges a FAPESP PhD fellowship no. 2018/22044-3. APV and SOS acknowledge the DGAPA–PAPIIT grant IA103224. SO acknowledges the support of the University of Padova, DOR Ortolani 2020, Piotto 2021 and Piotto 2022, Italy and funded by the European Union – NextGenerationEU" RRF M4C2 1.1  n: 2022HY2NSX. "CHRONOS: adjusting the clock(s) to unveil the CHRONO-chemo-dynamical Structure of the Galaxy” (PI: S. Cassisi). BB and EB acknowledge partial financial support from FAPESP, CNPq and CAPES - Financial code 001. B.D. acknowledges support by ANID-FONDECYT iniciación grant No. 11221366 and from the ANID Basal project FB210003. L.O.K. acknowledges partial financial support by CNPq (proc. 313843/2021-0) and UESC (proc. 073.6766.2019.0013905-48). This research is based on observations made with the NASA/ESA \emph{Hubble} Space Telescope obtained from the Space Telescope Science Institute, which is operated by the Association of Universities for Research in Astronomy, Inc., under NASA contract NAS 5–26555.  The HST observations are associated with programs GO-9799 (PI: Rich) and GO-15065 (PI: Cohen). Based on observations obtained at the international Gemini Observatory, a program of NSF NOIRLab, which is managed by the Association of Universities for Research in Astronomy (AURA) under a cooperative agreement with the U.S. National Science Foundation on behalf of the Gemini Observatory partnership: the U.S. National Science Foundation (United States), National Research Council (Canada), Agencia Nacional de Investigación y Desarrollo (Chile), Ministerio de Ciencia, Tecnología e Innovación (Argentina), Ministério da Ciência, Tecnologia, Inovações e Comunicações (Brazil), and Korea Astronomy and Space Science Institute (Republic of Korea). The Gemini observations are associated with the program GS-2020A-Q-121 (PI: L. Kerber).
\end{acknowledgements}

%
%

\bibliographystyle{aa} 
\bibliography{bibliog6558}

\begin{appendix} 

\section{Corner-plots}
\begin{figure}[hb!]
    \centering
    \includegraphics[width=0.49\textwidth]{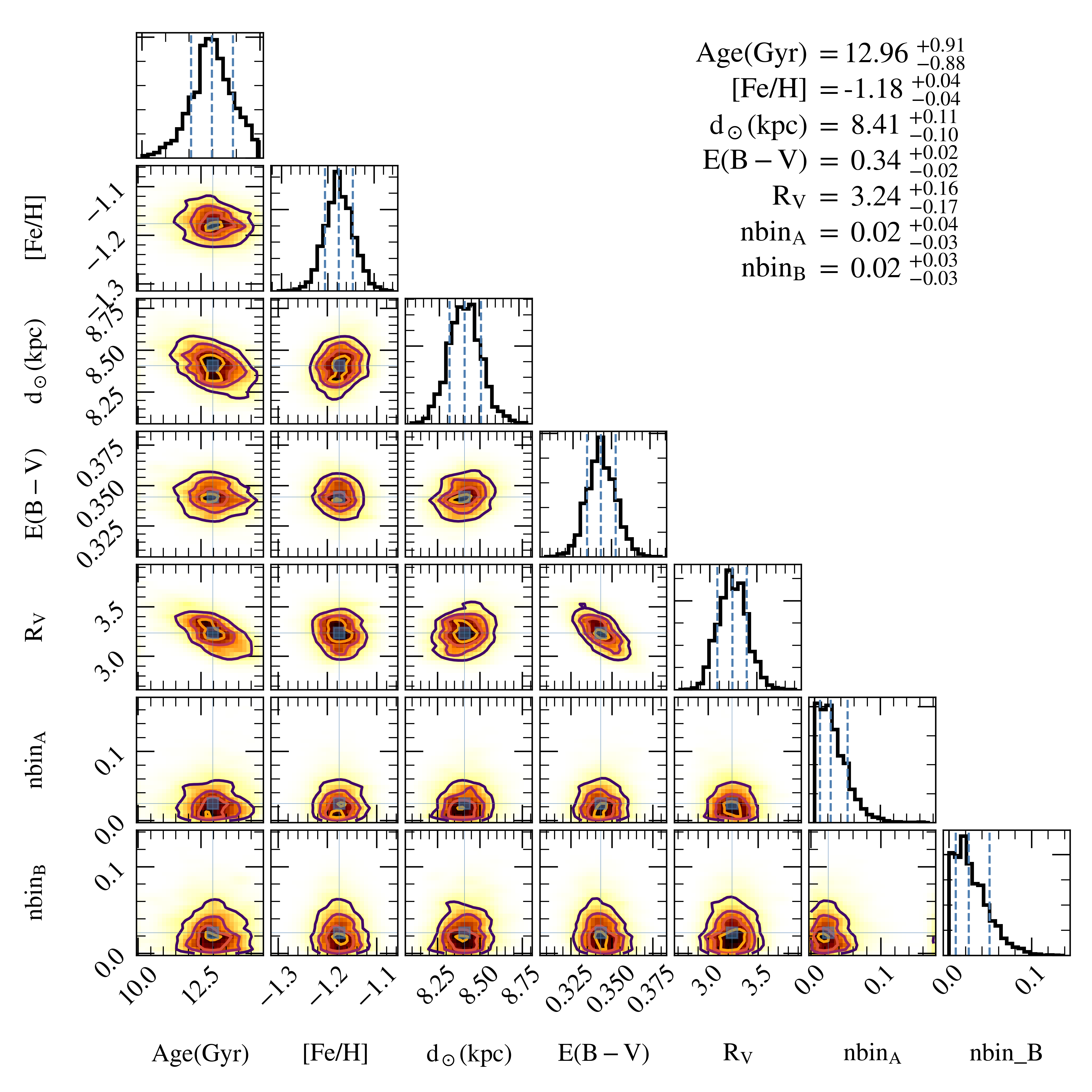}
    \includegraphics[width=0.49\textwidth]{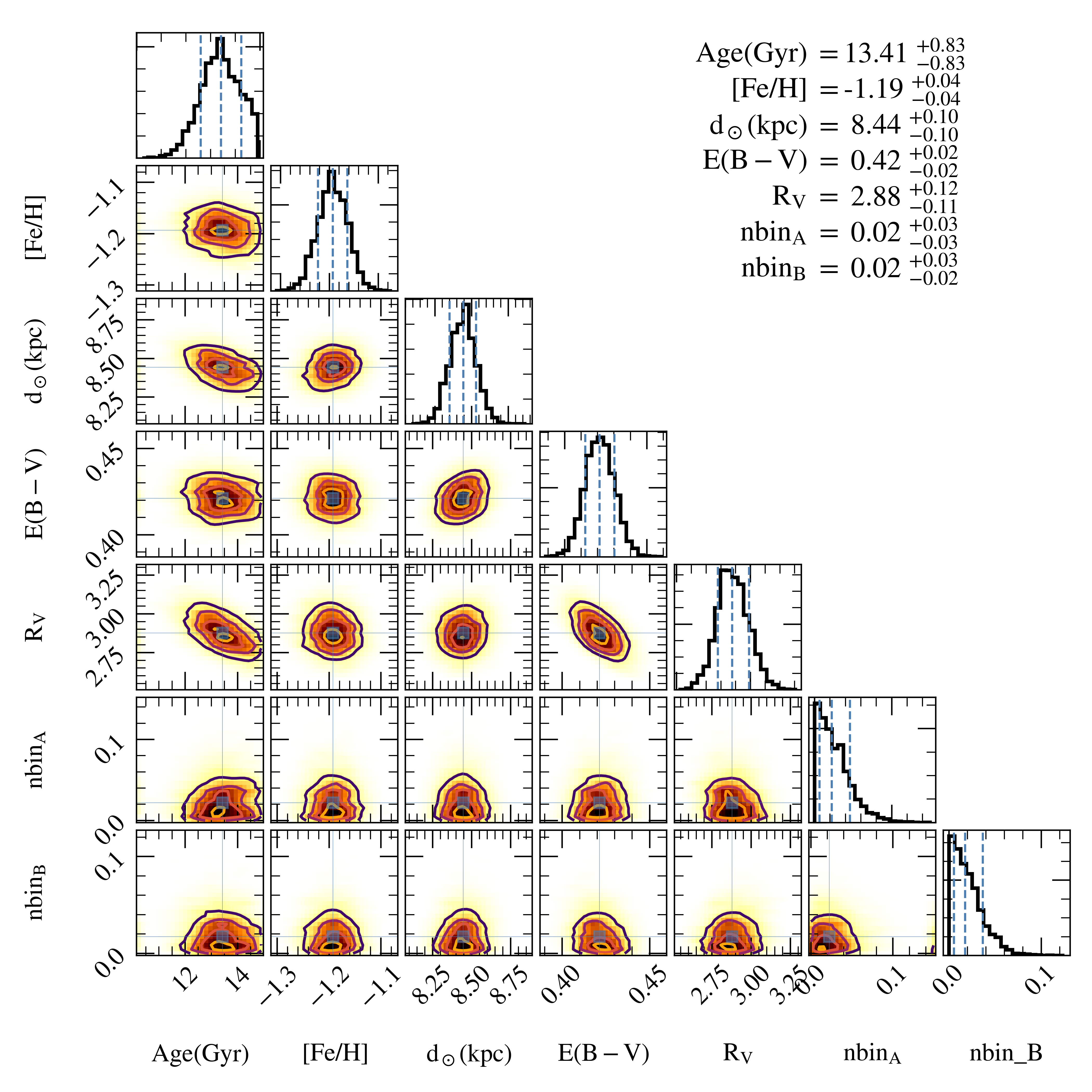}
    \caption{Corner-plots with the results for the isochrone fitting using DRCFL1 (upper panel) and DRCFL2 (bottom panel). The characteristic values are composed of the median and the 16 and 84 percentiles of the distributions.}
    \label{fig:corners}
\end{figure}

In this section we show the corner-plots for the simultaneous isochrone fitting. The binary fractions are different for each CMD: $n_a$ for the NIR and $n_b$ for the HST CMDs. In Figure \ref{fig:corners}, the upper panel shows the results for DRCFL1 and bottom panel for DRCFL2. The corner-plots are an important diagnostical too to visualize the behaviour of the McMC and the correlations among the parameter as well.

\end{appendix}

\end{document}